\documentclass[aps, prb, reprint, lengthcheck, footinbib, showpacs, superscriptaddress, citeautoscript]{revtex4-1}

\usepackage{amsmath}
\usepackage{amssymb}
\usepackage{braket}
\usepackage{graphicx}
\usepackage{euscript}
\usepackage[cmyk]{xcolor}
\usepackage[unicode, colorlinks, urlcolor=blue, citecolor=blue, linkcolor=blue]{hyperref}
\usepackage[all]{hypcap}
\usepackage[cp1251]{inputenc}
\usepackage[english]{babel}
\usepackage{cmap}
\usepackage{blindtext}

\input{tcilatex}
\begin{document}

\title{Effect of uniaxial stress on the interference of polaritonic waves
in wide quantum wells}
\author{D.~K.~Loginov}
\affiliation{St-Petersburg State University, St-Petersburg, 198504, Russia}
\author{A.~V.~Trifonov}
\affiliation{Spin Optics laboratory, St-Petersburg State University,
St-Petersburg, 198504, Russia}
\author{I.~V.~Ignatiev}
\affiliation{Spin Optics laboratory, St-Petersburg State University,
St-Petersburg, 198504, Russia}

\date{\today}

\begin{abstract}
A theory of polaritonic states is developed for a nanostructure with a wide quantum well stressed perpendicular to the growth axis of the heterostructure. The role of the $K$-linear terms appearing in the exciton Hamiltonian under the stress is discussed. Exciton reflectance spectra are theoretically modeled for the nanostructure. It is predicted that the spectral oscillations caused by interference of the exciton-like and photon-like polariton modes disappear with the increase of applied pressure and then appear again with opposite phase relative to that observed at low pressure. Effects of gyrotropy and  convergence of masses of excitons with heavy and light holes due to their mixing by the deformation is also considered. Numerical estimates performed for the GaAs wells show that these effects can be experimentally observed at pressure $P < 1$~GPa for the well widths of a fraction of micron.
\end{abstract}

\pacs{71.35.-y, 71.36.+c, 71.70.Fk, 78.67.De}

\maketitle




\section*{Introduction}

Interference of the polariton modes in thin crystals and heterostructures with wide quantum wells (QWs) is studied since 1973~\cite{Kiselev}. Because the QW is a resonator for polariton waves. Propagating across the QW layer, the interference gives rise to standing waves, which manifest themselves as oscillations in the reflection or absorption spectra. Such oscillations were really observed in the polariton spectra of the wide QWs and 
reasonable agreement between the theoretical calculations and experimental data was demonstrated~\cite{Kiselev, Kiselev2, Tomasini, Markov, Loginov}. The calculations were based on a simplest model, in which the exciton energy  quadratically depends on the exciton wave vector $\mathbf K$.
Such a model has limited applicability. In particular, in crystals with no inversion symmetry, the  kinetic energy of excitons can contain not only $\mathbf K$-quadratic but also $\mathbf K$-linear terms. The presence of linear terms in the dispersion low for polariton modes in a QW can lead to a number of new optical effects.

As it was shown in Refs.~\cite{Selkin, IvchSelk, AgranovichGinzburg, SirotinShaskolskaya}, the presence of $\mathbf K$-linear terms in the exciton Hamiltonian can result in a gyrotropy of the system. For the bulk crystals discussed in these papers, gyrotropy leads to the appearance of ellipticity in the reflected linearly polarized light. An example of calculations of the reflection coefficient for a crystal plate, taking into account the gyrotropy due to $\mathbf K$-linear terms is given in Ref.~\cite{AgranovichGinzburg}
. However, in oder to simplify the analysis the authors considered a semi-infinite gyrotropic crystal. The author of Ref. \cite{Golub}  discussed a reflection from the heterostructure with a QW in the spectral range of inter-subband electron transitions taking into account a natural optical activity described by the $\mathbf K$-linear terms.  The influence of $\mathbf K$-linear terms on the interference of polariton waves in wide QWs has not been studied so far.
 
The $\mathbf K$-linear terms can arise in the excitonic Hamiltonian due to the  $\mathbf k$-linear terms of the valence band Hamiltonian (with $\mathbf k$ denoting the wave vector of hole)~\cite{Kiselev, Kiselev2, Markov}.  However, their value is generally much less than that of the $\mathbf K$-quadratic terms (see, e.g.~\cite{PikusMaruschak}). 
For this reason, the model, which ignores the $\mathbf K$-linear terms, typically gives rise to reasonable agreement with the experiment as it was found in Refs. \cite{Kiselev, Kiselev2, Markov}. At the same time, the role of $\mathbf K$-linear terms in the crystals without inversion symmetry can be dramatically increased if an  uniaxial stress is applied to the cristall. The stress results in the  $\mathbf K$-linear splitting of the lowest conduction band and the upper valence band (see~\cite{PikusMaruschak, Cho}). Although the effect of uniaxial stress has been known for several decades~\cite{Cho}), its possible impact on the exciton and, accordingly, on the polariton spectra of wide QWs has not been investigated yet.

In this paper we present a theoretical analysis of interference of polariton waves in a wide GaAs QW subject to the uniaxial stress. We discuss the effects induced  by the stress applied perpendicular to the direction of the propagation of polariton waves. As shown below, the stress leads to linear in wave vector contributions to the energy of electrons and holes. This results in the modification of polaritonic states in the QW and, as a consequence, in the dramatic change of reflection spectra. In addition, the uniaxial stress leads to intermixing of the heavy-hole and light-hole excitons and to a ``convergence'' of their effective masses.

Analysis of the effect of uniaxial stress on the polariton states will be performed in the following sequence. In Section \ref{sec:Ham}, we consider the Hamiltonian of an exciton in the presence of the uniaxial stress without an interaction with light. Section \ref{sec:perm} is devoted to description of the permittivity of the strained crystal in the presence of the exciton-light interaction. Besides, calculations of the dispersion relations for the polariton modes are given in this section. In section \ref{sec:boundc}, we define the boundary conditions for the polariton modes. The main results are present in section \ref{sec:main}. The microscopic nature of the suppression and recovery of oscillations in polariton spectra is discussed in section \ref{sec:dissc}. The last section summarizes the main findings and conclusions. 

\section{Hamiltonian of an exciton in presence of stress}
\label{sec:Ham}

Consider an exciton in a crystal with the zinc blende symmetry, propagating along the $Z$ axis, which coincides with the [001] crystallographic axis. Such exciton is characterized by only one non-zero component of the exciton wave vector $K=K_z,~K_x=K_y=0$. In what follows, we consider this direction as the quantization axis for the angular momenta of carriers.  

The exciton states observed in the optical experiments in crystals like GaAs are formed from the states of the doubly degenerate conduction band  $\Gamma_6$ and the four-fold degenerate valence band $\Gamma_8$. The wave functions of electrons and holes in this approximation are, respectively, two- and four-component plane waves~\cite{LuttingerKhon,Luttinger}.

We derive the excitonic Hamiltonian from the Hamiltonians of free electrons and holes. In what follows, we neglect the terms related to the corrugation of the valence band~\cite{ChoSuga}, which are responsible for mixing heavy and light holes. These effects are weak compared to those caused by the terms included in the spherically-symmetric part of the valence band Hamiltonian and do not affect the phenomena discussed in the present paper.

To construct the excitonic Hamiltonian, we replace the coordinates of free electrons, $\mathbf r_e $, and of heavy and light holes, $\mathbf r_{hh}$, $\mathbf r_ {lh}$, by coordinates of motion of the exciton as a whole, $Z= (z_em_e + z_ {hh, lh} m_ {hh, lh}) / (m_e + m_ {hh, lh})$, and of the relative motion of the electron and hole, $\mathbf {r}=\mathbf r_e-\mathbf r_ {hh, lh}$. Here $m_e$, $m_ {hh}$, $m_ {lh}$  are effective masses of the electron, and of the heavy and light holes, respectively.
  
The effects considered below are observed in the spectral range where the kinetic energy of the exciton is comparable with its binding energy. Excitons with such kinetic energy can be described in the approximation of the ``large'' wave vector~\cite{ChoSuga}. According to this approximation, operators of the wave vectors for free electrons and holes can be expressed as follows:
\begin{equation}
\hat k^{(e,hh,lh)}_x=\pm\frac{\hat p_x}{\hbar},~\hat k^{(e,h)}_y=\pm\frac{\hat p_y}{\hbar},
\label{eq:tranzoper}
\end{equation}
$$\hat k^{(e,hh,lh)}_z=\pm\frac{\hat p_z}{\hbar}+\frac{m_{e,hh,lh}}{m_e+m_{hh,lh}}\hat K_z,$$
where signs ``+'' and ``-'' refer to electrons and holes, respectively. Here  $m_ {hh,lh} = m_0/ (\gamma_1 \pm \gamma_2) $, where $m_0$  is a free electrons mass and  $\gamma_1,~ \gamma_2$ are the Luttinger parameters \cite{Luttinger}. Operator $\hat p_ {x (y, z)} =-i \hbar \partial / \partial x (y, z) $  is the momentum operator of relative motion of the electron and hole;  $\hat K_z =-i\partial /\partial Z$ is the operator of wave vector of the exciton motion as a whole; $M_{h,l}=m_e+m_{hh,lh}$  denote translational masses of the heavy-hole and light-hole excitons. The Hamiltonian of the excitons is:
\begin{equation}
\hat H^{(0)}_{Xh,l}=\hat H^{(0)}_{Kh,l}+\hat H^{(0)}_{p},
\label{eq:exall}
\end{equation}
where
\begin{equation}
\hat H^{(0)}_{Kh,l}=E_g+\frac{\hbar^2\hat K^2}{2M_{h,l}}
\label{eq:exK}
\end{equation}
is the Hamiltonian of the exciton motion as a whole;
\begin{equation}
\hat H^{(0)}_{p}=\frac{\hat p^2}{2\mu}-\frac{e^2}{\epsilon_0r}
\label{eq:exrel}
\end{equation}
is the Hamiltonian of relative motion of the electron and hole. In Eqs. (\ref{eq:exall} - \ref{eq:exrel}), $E_g$ is a band gap, $\mu =1/m_e+\gamma_1/m_0$ is a reciprocal exciton mass,
$\epsilon_0$  is a background permittivity of the crystal, and $e$ is the electron charge. The coordinate part of the wave function of exciton has the form:
\begin{equation}
\Psi(K)=e^{iKZ}F_{{\cal NLM}},
\label{eq:vawefunckt}
\end{equation}
where $F_{{\cal NLM}}$ is the hydrogen-like wave function of the relative motion of  electron and hole.  This function can be represented as: $F_{{\cal NLM}}={\cal R}_{{\cal NL}}{\cal Y}_{{\cal LM}}$, where ${\cal R}_{{\cal NL}}$ and ${\cal Y}_{{\cal LM}}$ are radial and spherical functions, respectively.   Subscript ${\cal N}$  is the principal quantum number and subscripts ${\cal L},~{\cal M}$ are the orbital angular momentum and its projection on the quantization axis, respectively.

 The Hamiltonian Eq. (\ref{eq:exall}) does not mix orbital states of the exciton.  Uniaxial stress can lead to mixing of 1s- and p- exciton states, however, this effect can be neglected because of its smallness, (see below). Hence we consider only the 1s-exciton state with $N=1$, $L = 0$, and $M = 0$. Wave function of this state contains components: ${\cal R}_{10}=2e^{-r/a_B}$ and ${\cal Y}_{00}=1/\sqrt{4\pi}$,  where $a_B=\epsilon_0\hbar^2/\mu e^2$ is the exciton Bohr radius. Note that the coordinate parts of wave functions for the heavy-hole and light-hole excitons are the same up to some constant~\cite{Lipari1}. The energy of 1s-exciton in the unstrained crystal has the form:
\begin{equation}
H^{(0)}_{Xh,l}=E_X+\frac{\hbar^2 K^2}{2M_{h,l}},
\label{eq:zeroenergy}
\end{equation}
where $E_X=E_g-R$  is the energy of an optical transition to the exciton ground state, $R=\mu e^4/(2\hbar^2\epsilon_0^2)$ is the exciton binding energy, which is the eigenvalue of the operator Eq. (\ref{eq:exrel}). 

Hamiltonian of the exciton motion in the unstrained crystal can be written as a matrix $8\times8$ consisting of two identical blocks $4\times4$, which have the form:
 \begin{equation}
\hat H^{(0)}_X=\begin{pmatrix}
H^{(0)}_{Xh} & 0 & 0 & 0\\
0 & H^{(0)}_{Xl} & 0 & 0\\
0 & 0 & H^{(0)}_{Xl} & 0\\
0 & 0 & 0 & H^{(0)}_{Xh}
\end{pmatrix}.
\label{eq:zeromatrix}
\end{equation}
One of these blocks describes the optically active exciton states with spin projections $\pm1$, and the second one is the optically inactive states with spin projections $\pm2$ and $0$  for light-hole and heavy-hole excitons, respectively. The matrix of the Hamiltonian (\ref{eq:zeromatrix}) is written in the basis of 8-component plane waves, which we denote as:
\begin{equation}
|j,s\rangle_K=\nu_{j,s}\Psi(K).
\label{eq:wavepure}
\end{equation}
Here $\nu_{j, s}$ are the eight-component spinors, in which one component is equal to unit, and others are zero;  $j = \pm3 / 2, ~ \pm1/2 $ and $s =\pm1/2 $ are the projections of the spin moment of electron and hole, respectively, and   $\Psi(K) $  is given by the Eq.  (\ref{eq:vawefunckt}).

Uniaxial stress leads to two main effects. One of them is a change of the energy structure of the $\Gamma_8$ valence band. This effect is described by the Bir-Pikus Hamiltonian \cite{BirPikus} :
\begin{equation}
\hat H_\varepsilon=-a\mathbb I \mathrm{Sp}(\varepsilon)+b\sum_\alpha J_\alpha^2\Bigl(\varepsilon_{\alpha\alpha}-\frac13\mathrm{Sp}(\varepsilon)\Bigr)+
\label{eq:BirPikusfull}
\end{equation}
$$d\sum_{\alpha\ne\beta}\varepsilon_{\alpha\beta}\{J_\alpha,J_\beta\},$$
where $\mathbb I$ denotes the unit matrix, $\varepsilon_{\alpha\beta}$ is the strain tensor, $\mathrm{Sp}(\varepsilon)=\sum_{\alpha}\varepsilon_{\alpha\alpha}$. Matrices $J_\alpha$ denote the hole angular momentum, where $\alpha,~\beta=x,~y,~z$. Quantities $a$, $b$, and $d$ are the deformation potentials.

The second effect, considered here in more detail, is the appearance of $k$-linear terms  in the Hamiltonian of electron and holes for crystals without inversion symmetry ~\cite{PikusMaruschak, Cho}:
\begin{equation}
\hat H^{(\varepsilon k)}_c=\frac12(C_3\sum_\gamma \sigma_\gamma \hat \varphi_{\gamma}+C'_3\sum_\gamma \sigma_\gamma \hat \psi_\gamma),
\label{eq:linearelectron}
\end{equation}
\begin{equation}
\hat H^{(\varepsilon k)}_v=C_5\sum_\gamma J_\gamma \hat \varphi_\gamma+C_6\sum_\gamma J_\gamma \hat \psi_\gamma+C_7\sum_\alpha J_\gamma^3\hat \varphi_\gamma+
\label{eq:linearhole}
\end{equation}
$$C_8\sum_\gamma J_\gamma^3\hat \psi_\gamma+C_9\sum_\gamma V_\gamma\hat \chi_\gamma,$$
where $\gamma=x,~y,~z$; quantities  $\sigma_\gamma$ are the Pauli matrixes, $V_z=J_z(J_x^2-J_y^2)$, $C_j,~C'_j$ are material constants for a crystal under consideration ($j=3,~4,...9$). Components of operators $\hat \varphi_ \gamma $, $\hat\psi_\gamma$ and $\hat\chi_\gamma$ required for further consideration read:
$$\hat \varphi_z=\varepsilon_{xz}\hat k_x-\varepsilon_{yz}\hat k_y,$$
\begin{equation}
\hat \psi_z= \hat k_z(\varepsilon_{xx}-\varepsilon_{yy}),
\label{eq:hatpsiz}
\end{equation}
$$\hat \chi_z=\hat k_z (\varepsilon_{zz}-\frac13\mathrm{Sp} (\varepsilon)).$$
Other ($x$ and $y$) components of these operators can be obtained by the cyclic permutation of subscripts. 

Analysis shows, that, to describe the phenomena under discussion, one should consider only terms $C_6J_z\psi_z$, $C_7J_x^3\varphi_x$, $C_7J_y^3\varphi_y$, and $C_8J^3_z\psi_z$ [see Appendix A]. 

In what follows, we assume that pressure $P$ is applied along axis $x$,  which coincides with the $\textrm{C}_4$ ($[100]$) axis of crystal lattice. Under such conditions, all the off-diagonal components of the strain tensor are zero. The component of stress tensor $u_{xx}$ is equal in magnitude to the applied pressure $P$ and the diagonal components of strain tensor are described by  expressions~\cite{PollakCardona}:
$$\varepsilon_{xx}=S_{11}u_{xx}=S_{11}P,$$
\begin{equation}
\varepsilon_{yy}=S_{12}u_{xx}=S_{12}P,
\label{eq:epsilonP}
\end{equation}
$$\varepsilon_{zz}=S_{12}u_{xx}=S_{12}P,$$
where $S_{\alpha\beta}$ are the components of the elastic compliance tensor. 

Operators  $\varphi_x$ and $\varphi_y$ are zero because they contain off-diagonal components  $\varepsilon_{xz}$ and $\varepsilon_{yz}$ [see Eq. (\ref{eq:hatpsiz})], which are zero in the considered geometry. Therefore only two terms, $C_6J_z\psi_z$, and $C_8J^3_z\psi_z$, should be finally taken into account. Substitution of expressions (\ref{eq:tranzoper}) and (\ref{eq:epsilonP}) into Eq.~(\ref{eq:linearhole}) gives rise to the following expression for the stress-induced terms in excitonic Hamiltonian:
\begin{equation}
\hat H^{(K\varepsilon)}=\frac{m_h}{M}(C_6J_z+C_8 J^3_z)\Bigl(S_{11}-S_{12}\Bigr)P\hat K_z.
\label{eq:linear1}
\end{equation}
This operator calculated using wave functions  (\ref{eq:wavepure}) is the $8\times8$ diagonal matrix. Its nonzero matrix elements have the form:
\begin{equation}
H_{j+s,j+s}^{(K\varepsilon)}=A_{j} K,
\label{eq:exepsK}
\end{equation}
$$A_{j}\equiv\frac{m_h}{M}\Bigl(jC_6+j^3C_8\Bigr)\Bigl(S_{11}-S_{12}\Bigr)P,$$
where  $j=\pm3/2$,$\pm1/2$ and $s =\pm1/2$. Note that the sign of constant
 $A_{j}$ is determined by the sign of the angular momentum projection $j$ for the  hole so that  $A_{j}=-A_{-j}$. Parameters  $C_6$ and  $C_8$ can be determined using expressions given in Ref.~\cite{PikusMaruschak}. Analysis shows that $A_{\pm1/2}=0$ for the light-hole exciton in all crystals with zinc-blende structure. Quantity $A_{\pm3/2}$  describing the effect of $K$-linear terms for the heavy-hole exciton is completely determined by the applied stress
and by material parameters. Its value for the QW under consideration is given in Sect. \ref{sec:perm}.

Let us now consider in more detail the effects described by the Bir-Pikus Hamiltonian (\ref{eq:BirPikusfull}). At the chosen direction [100] of applied stress, this Hamiltonian reads:
\begin{equation}
\hat H_\varepsilon=-a\mathbb I \mathrm{Sp}(\varepsilon)+b\sum_\alpha J^2_\alpha\Bigl(\varepsilon_{\alpha\alpha}-\frac13\mathrm{Sp}(\varepsilon)\Bigr),
\label{eq:BirPikus}
\end{equation}
where $\varepsilon_{xx},~\varepsilon_{yy},~\varepsilon_{zz}$ are described by expressions  (\ref{eq:epsilonP}).

In addition to the shift of valence band, this Hamiltonian describes mixing of heavy and light holes, since matrices $J_x^2$ and $J_y^2$  contain off-diagonal elements. This mixing is significant because deformation potentials $a$ and $b$ are large. In particular, they are in orders of magnitude larger than the matrix elements of Hamiltonian (\ref{eq:linear1}) for the actual range of wave vectors $K$  ($0 \div$ $5\cdot10 ^ {6} $ cm$^{-1}$).

Finally, the total excitonic Hamiltonian in the presence of uniaxial stress consists of Hamiltonians  (\ref{eq:zeromatrix}), (\ref{eq:linear1}) and (\ref{eq:BirPikus}). This Hamiltonian does not mix optically active states, $ |j, s\rangle = |\pm 3/2,\mp 1/2\rangle,~|\pm 1/2,\pm 1/ 2\rangle$, with optically inactive states, $ |j, s\rangle = |\pm 3/2,\pm 1/2\rangle,~|\pm 1/2,\mp 1/ 2\rangle$, as it follows from properties of matrices $J_z,~J_z^3$, $J_x^ 2,~J_y^2,~J_z^2$  (see, e.g.,~\cite{Luttinger}). Therefore, we further restrict our analysis only by the bright excitons. Matrix of the total exciton Hamiltonian built on the wave functions of the bright exciton has the form:
\begin{equation}
\hat H_{X}=\hat H_{X}^{(0)}+ \hat H^{(K\varepsilon)}+\hat H_{\varepsilon}=\begin{pmatrix}
H_{h+} & 0 & V & 0\\
0 & H_{l+} & 0 & V\\
V & 0 & H_{l-} & 0\\
0 & V & 0 & H_{h-}
\end{pmatrix},
\label{eq:hamexfull}
\end{equation}
where 
$$H_{h\pm}=H^{(0)}_{Xh}+H_{\varepsilon h}\pm A_{3/2}K,$$ $$H_{l\pm}=H^{(0)}_{Xl}+H_{\varepsilon l},$$
$$V=\frac32b(S_{11}-S_{12})P,$$
$$H_{\varepsilon h,l}=-a\textrm{Sp}(\varepsilon)\pm\frac b2(\varepsilon_{xx}+\varepsilon_{yy}-2\varepsilon_{zz}).$$
In the last expression, the upper signs refer to the heavy-hole while the lower ones, to light-hole excitons.

The exciton wave function for this problem is constructed as a linear combination of  the basic wave functions (\ref{eq:wavepure}):
\begin{equation}
\Psi(K)=\sum_{j,s}C_{j,s}|j,s\rangle,
\label{eq:wavemixex1}
\end{equation}
where $C_{j,s}$ are the expansion coefficients.

\section{Permittivity tensor}
\label{sec:perm}

For calculation of the reflection spectra in the presence of the pressure-induced effects, we use the model of interference of the bulk polariton waves described in Refs.~\cite{Kiselev, Kiselev2, Tomasini, Markov, Loginov} . We consider the model of a heterostructure consisting of the QW layer surrounded by semi-infinite barriers. 
We assume that the incident light propagates perpendicularly to the QW and has a circular polarization. The latter corresponds to the creation of an exciton with a certain projection of the angular momentum on the direction of propagation.

For further analysis, we consider the wave to be right-hand polarized if the projection of the photon angular momentum  onto the chosen axis $Z$ is $+1$ 
and to be left-hand polarized if this projection is $-1$. With this definition, the sign of the circular polarization is not changed when the propagation of light is changed to opposite direction. Such a definition is convenient when writing the boundary conditions for exciton polaritons, see next section. It should be emphasized that this definition does not match with the commonly used one for the polarization, which is determined by the angular momentum projection of photons 
on the direction of propagation and, therefore, is reversed under reflection.

As the first step, we should calculate the permittivity tensor of the medium, $\epsilon (\omega,K)$, taking into account the exciton-photon interaction. Tensor   $\epsilon(\omega,K)$ is the $3\times3$ matrix describing two transverse and one longitudinal modes. However, under normal incidence of light, only transverse modes are excited in the crystal, and the $2\times2$ permittivity tensor is relevant to this case. 

The exciton-photon interaction is described by the perturbation operator (\cite{Pekar, Chopolarit, NozueCho}):
\begin{equation}
(d_h^{(\pm)}+d_l^{(\pm)})E^{(\pm)},
\label{eq:photex}
\end{equation}
where $E^{(\pm)}$ is the electric field amplitude of the light wave and superscript ``$\pm$'' corresponds to two circular polarizations of light.
The matrix element of the dipole moment operator $\hat d_{\pm}=e(x \pm y)$, has the form (\cite{Pekar, Chopolarit, NozueCho}):
$$d^{(\pm)}_h=\langle 0|\hat d^{(\pm)}|\pm 3/2,\mp1/2\rangle,~d^{(\pm)}_l=\langle 0|\hat d^{(\pm)}|\pm 1/2,\pm1/2\rangle.$$
Wave function $|0\rangle$ describes the vacuum state of crystal that is the state with no exciton. For both circular polarizations,  $| d_ {h, l} ^ {(+)} | = | d_ {h, l} ^ {(-)} | = d_ {h, l} $, where $d^2_h=3d^2_l=\frac34\hbar\omega_ {LT}\epsilon_0$ is the squared matrix element of the dipole moment normalized to unit volume. Quantity  $\hbar\omega_{LT}$ is the energy of longitudinal-transverse splitting describing the strength of the exciton-photon interaction.

The exciton-photon interaction changes polarization of the medium, which is described by expression \cite{Pekar, Chopolarit, NozueCho}:
\begin{equation}
{\cal P}^{(\pm)}=d_hC_{\pm3/2,\mp1/2}+d_lC_{\pm1/2,\pm1/2},
\label{eq:pertexmed}
\end{equation}
where $C_{js}$ are the expansion coefficients in expression~(\ref{eq:wavemixex1}). Polarization vector ${\cal P}$ are associated with vector $E^{(\pm)}$ via electric susceptibility tensor $4\pi\chi_{\alpha\beta}$:
$${\cal P}^{(\pm)}=4\pi\chi_{\pm\pm}E^{(\pm)}+4\pi\chi_{\pm\mp}E^{(\mp)}.$$

The dispersion relations, the wave functions and the permittivity tensor of the medium in the presence of the exciton-photon interaction can be found using the method proposed in Refs. \cite{Pekar, Chopolarit, NozueCho}. To this end, one should, first, solve a system of equations for the energies of the four states with perturbation   (\ref{eq:photex}) (see, e.g., \cite{Chopolarit, NozueCho}):
\begin{equation}
(\mathbb H-\mathbb I\hbar\omega)\mathbb C=0,
\label{eq:matrixes}
\end{equation}
where $\hbar\omega$ is the photon energy and matrix $(\mathbb H-\mathbb I\hbar\omega)$ and vector $\mathbb C$ have the form:
\begin{widetext}
\begin{equation}
(\mathbb H-\mathbb I\hbar\omega)=
\begin{pmatrix}
H_{h+}-\hbar\omega & 0 & V & 0 & d_h & 0 \\
0 & H_{l+}-\hbar\omega & 0 & V & d_l & 0 \\
V & 0 & H_{l-}-\hbar\omega & 0 & 0 & d_l\\
0 & V & 0 & H_{h-}-\hbar\omega & 0 & d_h \\
\end{pmatrix} 
,\text{ } \mathbb C=\begin{pmatrix}
C_{+3/2,-1/2}\\
C_{+1/2,+1/2} \\
C_{-1/2,-1/2}\\
C_{-3/2,+1/2}\\
-E^{(+)} \\
-E^{(-)}\\
\end{pmatrix}.
\label{eq:dispdet}
\end{equation}
\end{widetext}
Coefficients $C_{js}$ can be found solving system (\ref{eq:dispdet}):
\begin{equation}
C_{\pm3/2,\mp1/2}=\frac{\tilde H_{l\mp}d_hE^{(\pm)}}{\tilde H_{h\pm}\tilde H_{l\mp}-V^2}-\frac{Vd_lE^{(\mp)}}{\tilde H_{l\mp}\tilde H_{h\pm}-V^2},
\label{eq:wavemixcoef1}
\end{equation}
\begin{equation}
C_{\pm1/2,\pm1/2}=\frac{\tilde H_{h\mp}d_lE^{(\pm)}}{\tilde H_{l\pm}\tilde H_{h\mp}-V^2}-\frac{Vd_hE^{(\mp)}}{\tilde H_{l\pm}\tilde H_{h\mp}-V^2}.
\label{eq:wavemixcoef2}
\end{equation}
Here  $\tilde H_\alpha\equiv H_\alpha-\hbar\omega+i\Gamma$, where $\Gamma$ is a phenomenological parameter introduced to describe the processes of energy dissipation. Substituting (\ref{eq:wavemixcoef1}) and (\ref{eq:wavemixcoef2}) into equation (\ref{eq:pertexmed}), we obtain:
\begin{equation}
\frac{\tilde H_{l\mp}d^2_hE^{(\pm)}}{\tilde H_{h\pm}\tilde H_{l\mp}-V^2}+\frac{\tilde H_{h\mp}d^2_lE^{(\pm)}}{\tilde H_{l\pm}\tilde H_{h\mp}-V^2}-\frac{Vd_ld_hE^{(\mp)}}{\tilde H_{l\mp}\tilde H_{h\pm}-V^2}-
\label{eq:eq:56/6}
\end{equation}
$$\frac{Vd_ld_hE^{(\mp)}}{\tilde H_{l\pm}\tilde H_{h\mp}-V^2}=4\pi\chi_{\pm\pm}E^{(\pm)}+4\pi\chi_{\pm\mp}E^{(\mp)}$$
This relation allows one to finally obtain relations for components  $\chi_{\alpha,\beta}$:
\begin{equation}
4\pi\chi_{\pm\pm}=\frac{\tilde H_{l\mp}d^2_h}{\tilde H_{h\pm}\tilde H_{l\mp}-V^2}+\frac{\tilde H_{h\mp}d_l^2}{\tilde H_{l\pm}\tilde H_{h\mp}-V^2},
\label{eq:Pol/pm/pm}
\end{equation}
and
\begin{equation}
4\pi\chi_{+-}=4\pi\chi_{-+}=-\frac{Vd_ld_h}{\tilde H_{l\mp}\tilde H_{h\pm}-V^2}-\frac{Vd_hd_l}{\tilde H_{l\pm}\tilde H_{h\mp}-V^2}.
\label{eq:Pol/pm/mp}
\end{equation}

Beside the effects related to excitons, the uniaxial stress leads to a piezo-optical effect, due to which the isotropic crystal becomes uniaxial. In our case, the principal optical axis  of the crystal is perpendicular to the direction of light propagation. In the basis of circularly polarized light waves, this effect can be described by additional diagonal and off-diagonal terms in the permittivity tensor, which have the form \cite{etc1, etc2}:
\begin{eqnarray}
\delta\epsilon_{++}=\delta\epsilon_{--}=(\pi_{11}+\pi_{12}){ P}/2,\nonumber\\
\delta\epsilon_{+-}=\delta\epsilon_{-+}=(\pi_{11}-\pi_{12}){ P}/2.
\label{eq:eo^p}
\end{eqnarray}
Here $\pi_{11},~\pi_{12}$ are components of the background piezooptic tensor, whose values can be found in literature (see, e.g., \cite{etc1, etc2}).

The total permittivity is expressed in terms of the electric susceptibility (\ref{eq:Pol/pm/pm}) and (\ref{eq:Pol/pm/mp}) as follows:
\begin{equation}
\epsilon_{\pm\pm}(\omega,K)=\epsilon_0+\delta\epsilon_{++}+4\pi\chi_{\pm\pm},
\label{eq:circulmatrel}
\end{equation}
\begin{equation}
\epsilon_{+-}(\omega,K)=\epsilon_{-+}(\omega,K)=\delta\epsilon_{+-}+4\pi\chi_{+-}.
\label{eq:melnondiog}
\end{equation}
Permittivity tensor in the basis of circular polarizations has the form:
\begin{equation}
\epsilon(\omega,~K)=\begin{pmatrix}
\epsilon_{++} & \epsilon_{+-}\\
\epsilon_{-+} & \epsilon_{--}\\
\end{pmatrix}.
\label{eq:dielfun}
\end{equation}
It is easy to verify that, for the diagonal matrix elements of the tensor, the following relation is valid:
\begin{equation}
\epsilon_{++}(\omega,K)=\epsilon_{--}(\omega,-K).
\label{eq:diogonalsootncirc}
\end{equation}
This relation is equivalent to the a well-known relation \cite{AgranovichGinzburg}:
\begin{equation}
\epsilon_{xy}(\omega,K)=\epsilon_{yx}(\omega,-K),
\label{eq:Onzager}
\end{equation}
where $\epsilon_{xy},~\epsilon_{yx}$ are components of the permittivity tensor in the basis of linearly polarized waves. They are related to components (\ref{eq:circulmatrel}) and (\ref{eq:melnondiog}) by well known formulas (see, e.g., \cite{SirotinShaskolskaya})
:
\begin{equation}
\epsilon_{xy}(\omega,~K)=-\epsilon_{yx}(\omega,~K)=i\frac12(\epsilon_{++}(\omega,~K)-\epsilon_{--}(\omega,~K)).
\label{eq:dielfunlin}
\end{equation}
%

\begin{figure*}[t]
\includegraphics[clip,width=.99\columnwidth]{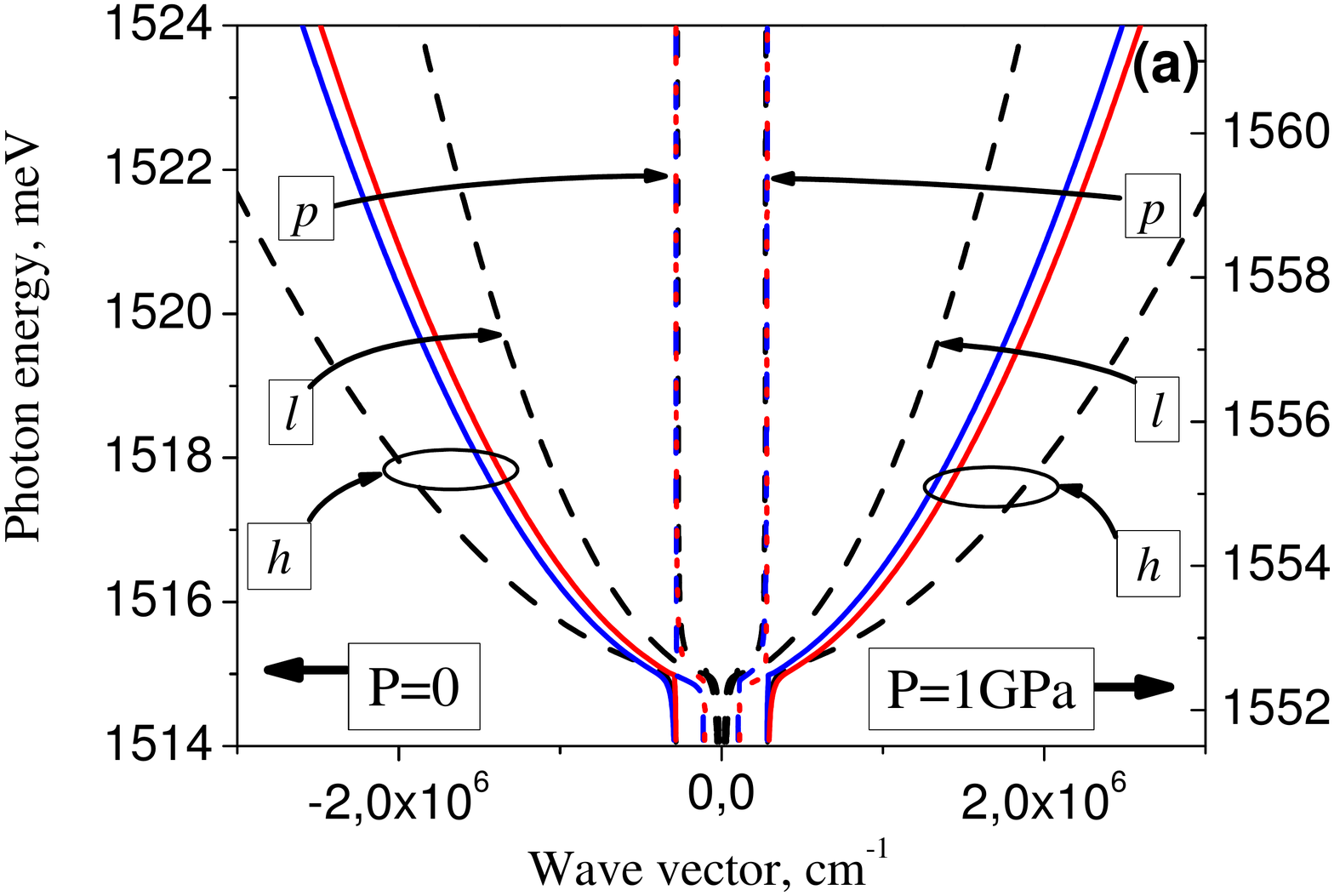}
\includegraphics[clip,width=.99\columnwidth]{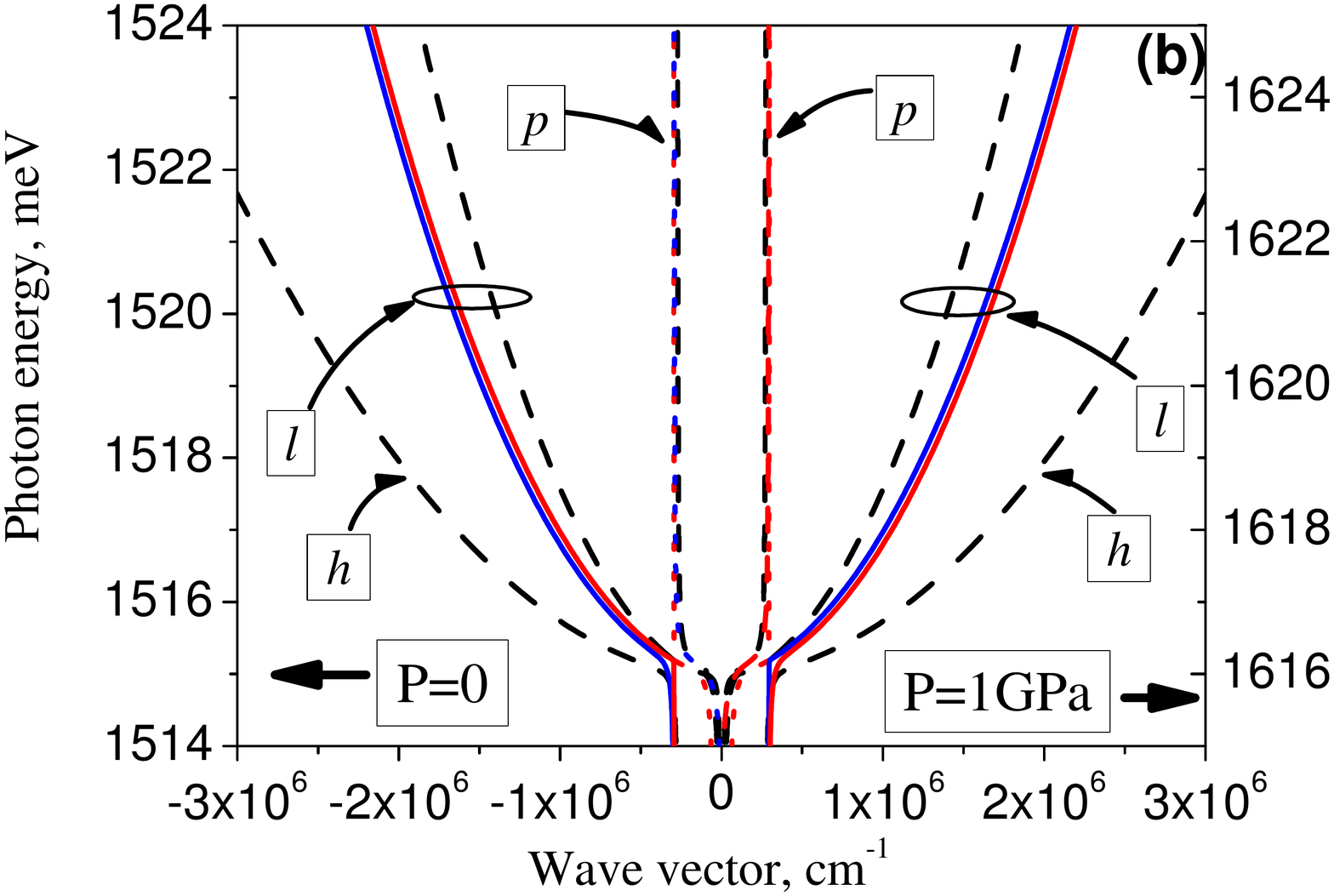}
\caption{The change of the dispersion relations of $h$- and $l$-polaritons branches  (figures (\textbf{a}) and (\textbf{b}), respectively) under pressure $P=1$ GPa.
Black dashed lines are the initial dispersion curves at $P=0$. Red and blue lines represent polariton modes with dominant right-hand and left-hand circularly polarized components in the stressed crystal, respectively. Legends $h$, $l$ and $p$ are explained in the text. The left and right energy scales correspond to cases $P=0$ and $P=1$ GPa. \vspace{0.5cm}}
\label{fig:1}
\end{figure*}

First, we use resulting expression (\ref{eq:dielfun}) for calculation of the dispersion relations for polariton eigen modes. To this end, we solve the dispersion equation \cite{AgranovichGinzburg}
:
\begin{equation}
\epsilon(\omega,K)=\mathbb I\frac{c^2K^2}{\omega^2},
\label{eq:disequat}
\end{equation}
where $c$ is the light velocity and $\epsilon(\omega,K)$ is the permittivity described by expressions (\ref{eq:circulmatrel}-\ref{eq:dielfun}). 

Equation (\ref{eq:disequat}) has 12 independent solutions for dispersion relations $K_i(\omega)$, half of which corresponds to the propagation of polariton waves in the forward direction and the other half, in the backward one. The eigenmodes differ from each other by the predominant contribution of the photon-like ($p$ - type) or exciton-like ($l$- and $h$- type) components, see Fig.~\ref{fig:1}. Since permittivity tensor (\ref{eq:dielfun}) has non-zero off-diagonal matrix elements, all the eigenmodes propagating in the strained crystal, in general, are elliptically polarized. In particular, six waves are predominantly left-hand polarized while other six are right-hand polarized.

Figure \ref{fig:1} shows the dispersion curves for polariton eigenmodes calculated for a GaAs crystal. The following material parameters are used: $\hbar\omega_{LT}=0.09$ meV (\cite{Schultheis}), $\epsilon_0=12.56$ (\cite{Stillman}), $m_e=0.067~m_0$ (\cite{Lawaetz}), $m_{hh}=0.45~m_0$, $m_{lh}=0.082~m_0$ (\cite{Skolnick}) ($m_0$ is the free electron mass), $E_g=1520$ meV (\cite{Sturge}), $R=5$ meV, $S_{11}=1.76\cdot 10^{-12}~\mathrm{cm^2/dyn}$ and $S_{12}=-0.37\cdot 10^{-12}~\mathrm{cm^2/dyn}$ \cite{Averkiev}, $\pi_{11}=0.2\text{ GPa}^{-1},~\pi_{12}=0.05\text{ GPa}^{-1}$,
$\pi_{11}-\pi_{12}\approx0.2\text{ GPa}^{-1}$ \cite{etc1, etc2} and $a=-6.7$ eV, $b=-1.7$ eV \cite{IvchenkoPikus}. Constants $C_6$ and $C_8$  are calculated using formulas and material constants given in Ref. \cite{PikusMaruschak}: $C_6=-1\text{ meV/cm}$ and $C_8=-4\text{ meV/cm}$. The value of the damping parameter has been chosen to obtain the width of oscillations in the calculated reflection spectra (see below) approximately equal to that typically observed in the experiment: $\Gamma=0.05$ meV. 
The figure shows the dispersion curves for pressure $P=0$ and $P=1$ GPa. Since the GaAs crystal is destroyed at larger uniaxial stress \cite{Ansp, CardonaPSS}, the pressure effects at $P>1$ GPa was not considered.

As seen Fig.\ref{fig:1} the strain results in two main effects. The first one is the change of curvature of the exciton-like dispersion branches. It is caused by mixing of the heavy-hole and light-hole exciton states and is described by the off-diagonal matrix element $V$ in matrix (\ref{eq:dispdet}).
Dispersion curves for $h$-type waves, which correspond to the heavy-hole excitons in the absence of strain, become steeper with increasing pressure [see Fig.\ref{fig:1}(a)] and those for the $l$-type waves initially corresponding to light-hole excitons become flatter [Fig.\ref{fig:1}(b)].
This effect can be treated as the convergence of the effective masses of excitons of $h$- and $l$-types. 

The second effect is the anti-symmetric in $K$ splitting of dispersion branches of $h$- and $l$-types, which is described by term  (\ref{eq:exepsK}) in the exciton Hamiltonian. 
If $K>0$, the dispersion branch for the right-hand polarization is higher in energy than the branch for the left-hand polarized component. If $K<0$, these branches are swapped (see Fig. \ref{fig:1}a). The relatively weak, at first glance, splitting of the dispersion curves leads to a qualitatively new effect in the reflection spectra of the QW.

\section{Boundary conditions}
\label{sec:boundc}

To formulate the boundary conditions, one should obtain a relation between the electric fields $E_\rho^{(+)}$ and $E_\rho^{(-)}$ for polariton  eigenmodes.The relation follows from Eqs. (\ref{eq:dielfun}) and (\ref{eq:disequat}):
\begin{equation}
 \mathbb I\frac{c^2K_\rho^2}{\omega^2}\mathbb E_\rho=\epsilon(\omega, K_\rho)\mathbb E_\rho,
\label{eq:sigpmmtr}
\end{equation}
where the vector
$$\mathbb E_\rho=\begin{pmatrix}
E_\rho^{(+)}\\
E_\rho^{(-)}
\end{pmatrix}$$
describes the elliptical polarization of the eigen mode $\rho$. Subscript $\rho$ iincludes three components:
$$\rho=\{\lambda,d,e\},$$
where  $\lambda=p,~l,~h$ indicates the type of the eigen mode,  $d=\rightarrow$ or $\leftarrow$ shows the direction of wave propagation. Index ``$e$'' specifies the ellipticity, in particular,  $e=right $ denotes the wave with prevailing right-hand  circularly polarized component, and $e=left$ is the wave with prevailing left-hand polarized component [see Fig.\ref{fig:1}].

Substitution of vector $\mathbb E_\rho$ in expression (\ref{eq:sigpmmtr}) provides relation between the circularly polarized components:
\begin{equation}
 \frac{c^2K_\rho^2}{\omega^2}E^{(\pm)}= \epsilon_{\pm\pm} E_\rho^{(\pm)}+\epsilon_{+-} E_\rho^{(\mp)},
\label{eq:conertfield}
\end{equation}
which can be written as:
\begin{equation}
E_\rho^{(\pm)}= \xi_\rho^{(\mp)} E_\rho^{(\mp)},
\label{eq:plmncoupl}
\end{equation}
where
$$\xi_\rho^{(\pm)}=\frac{\epsilon_{+-}(\omega,K_\rho)}{\epsilon_{\pm\pm}(\omega,K_\rho)-n_\rho^2(\omega,K_\rho)}.$$

\begin{figure}[t] 
\includegraphics[clip,width=.6\columnwidth]{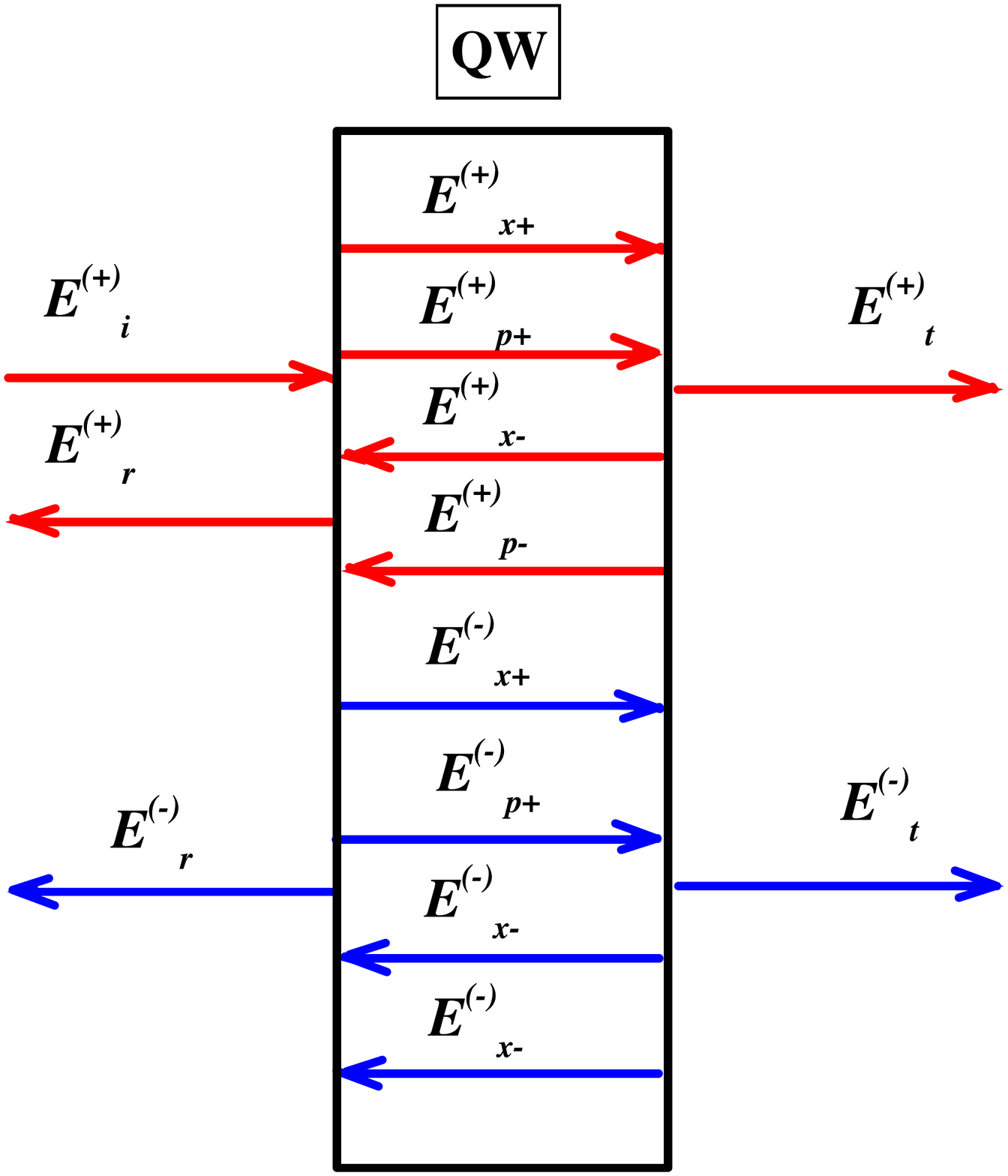}
\caption{Polaritonic eigenmodes in the QW layer. Red dashed arrows indicate polariton waves with predominant $\sigma^+$-polarization (subscript $right$). Blue dashed arrows show the polariton waves with a predominant $\sigma^-$ - polarization (subscript $left$). The solid red and blue arrows indicate the light waves in co- and cross-polarizations, respectively, passed through and reflected from the QW.
\vspace{0.5cm}}
\label{fig:structure}
\end{figure}

The polariton  eigenmodes modes propagating in the optically uniaxial QW are shown in Fig. \ref{fig:structure}. The circularly polarized incident light can excites all modes, but with different efficiency. The reflected light, in general, is elliptically polarized, i.e., can be decomposed into two circularly polarized components. Therefore, two sets of boundary conditions should be considered for each heterointerface, one set per each circular polarization. In what follows, we assume that the incident light has the right-hand helicity. 

Boundary conditions include the Maxwell's conditions, which require continuity of the tangential components of electric $E^{(\pm)}$ and magnetic $B^{(\pm)}$ fields of polaritonic waves at the QW heterointerfaces. For plane waves, the magnetic induction can be expressed in terms of the electric field amplitude: $B_\rho^{(\pm)}=n_\rho E_\rho^{(\pm)}$, where refractive index  $n_\rho=cK_\rho/\omega$. Thus, if there are incident, transmitted, and reflected waves at the QW heterointerfaces, the boundary conditions for the circularly polarized components can be written as:
\begin{equation}
E^{(+)}_i e^{iqZ_{1,2}}+E^{(+)}_g e^{-iqZ_{1,2}}=\sum_\rho E^{(+)}_\rho e^{iK_\rho Z_{1,2}}
\label{eq:Maxwellplus}
\end{equation}
$$n_0E^{(+)}_i e^{iqZ_{1,2}}-n_0E^{(+)}_g e^{-iqZ_{1,2}}=\sum_\rho n_\rho E^{(+)}_\rho e^{iK_\rho Z_{1,2}}
,$$
and
\begin{equation}
E^{(-)}_g e^{-iqZ_{1,2}}=
\sum_\rho E^{(-)}_\rho e^{iK_\rho Z_{1,2}},
\label{eq:Maxwellminus}
\end{equation}
$$-n_0E^{(-)}_g e^{iqZ_{1,2}}=\sum_\rho n_\rho E^{(-)}_\rho e^{iK_\rho Z_{1,2}} ,$$
where $E^{(+)}_i,~E^{(\pm)}_r$ are the amplitudes of the circularly polarized components of the incident, reflected ($g=r$), or transmitted ($g=t$) light waves outside the QW. Quantities $n_0$ and $q$ are the refractive index and the modulus of the wave vector of light in barriers, respectively. Coordinates $Z_{1}=0$ and $Z_{2}=L_{QW}$ correspond to the boundaries of the QW. 

Besides the Maxwell's boundary conditions , we use the Pekar's additional boundary conditions (ABC) According to them, the polarization of crystal caused by the heave-hole and light-hole excitons vanishes at the boundaries of the QW (see Ref. \cite{Pekar}). 
Since their contribution is described by expression (\ref{eq:pertexmed}), the ABC can be written as:
\begin{equation}
\sum_{\rho}d_hC_{\pm3/2,\mp1/2}(K_\rho)|_{L=0,~L_{QW}}=0,
\label{eq:vklad}
\end{equation}
$$\sum_{\rho}d_lC_{\pm1/2,\pm1/2}(K_\rho)|_{L=0,~L_{QW}}=0,$$
where $\rho$ runs over all polaritonic modes. Using expressions (\ref{eq:wavemixcoef1}) and (\ref{eq:wavemixcoef2}) for coefficients  $C_{\pm3/2,\mp1/2}$ and $C_{\pm1/2,\pm1/2}$ we obtain four ABC at each boundary:
\begin{equation}
\sum_{\rho} \frac{d^2_h\tilde H_{l\mp}(K_\rho) E^{(\pm)}_{\rho} e^{iK_{\rho} Z_{1,2}}}{\tilde H_{h\pm}(K_\rho)\tilde H_{l\mp}(K_\rho)-V^2}-\sum_{\rho} \frac{d_ld_hV E^{(\mp)}_{\rho} e^{iK_{\rho} Z_{1,2}}}{\tilde H_{h\pm}(K_\rho)\tilde H_{l\mp}(K_\rho)-V^2}=0,
\label{eq:Pekar}
\end{equation}
\begin{equation}
\sum_{\rho} \frac{d^2_l\tilde H_{h\mp}(K_\rho) E^{(\pm)}_{\rho} e^{iK_{\rho} Z_{1,2}}}{\tilde H_{l\pm}(K_\rho)\tilde H_{h\mp}(K_\rho)-V^2}-\sum_{\rho} \frac{d_hd_lV E^{(\mp)}_{\rho} e^{iK_{\rho} Z_{1,2}}}{\tilde H_{l\pm}(K_\rho)\tilde H_{h\mp}(K_\rho)-V^2}=0.
\label{eq:PekarL}
\end{equation}
It is easy to see that, at zero pressure when $V = 0$, these ABC are transformed into the ordinary Pekar's ABC.

\section{Reflectance spectra}
\label{sec:main}

Boundary conditions (\ref{eq:Maxwellplus}) - (\ref{eq:PekarL}) comprise a system of linear equations for amplitudes of the electric field of light waves and of polariton waves in the structure. Solution of this system allows one to determine amplitudes of the incident and reflected light waves in two polarizations and to calculate reflection coefficients:
$$R^{(++)}(\omega)=\frac{|E^{(+)}_i|^2}{|E_r^{(+)}|^2},$$
$$R^{(-+)}(\omega)=\frac{|E^{(-)}_i|^2}{|E_r^{(+)}|^2},$$
where superscripts ``$++$'' and ``$-+$'' denote the reflection coefficients in the co and cross polarizations.
We have carried out calculations of reflection spectra for the GaAs QW with thickness $L_{QW}=700$ nm. Background permittivity of the left and right semi-infinite space were chosen $\epsilon_l=1$ and $\epsilon_r=11$, which correspond to the air permittivity on the left side and to a typical semiconductor one on the right side.

\begin{figure}[t] 
\includegraphics[clip,width=.49\columnwidth]{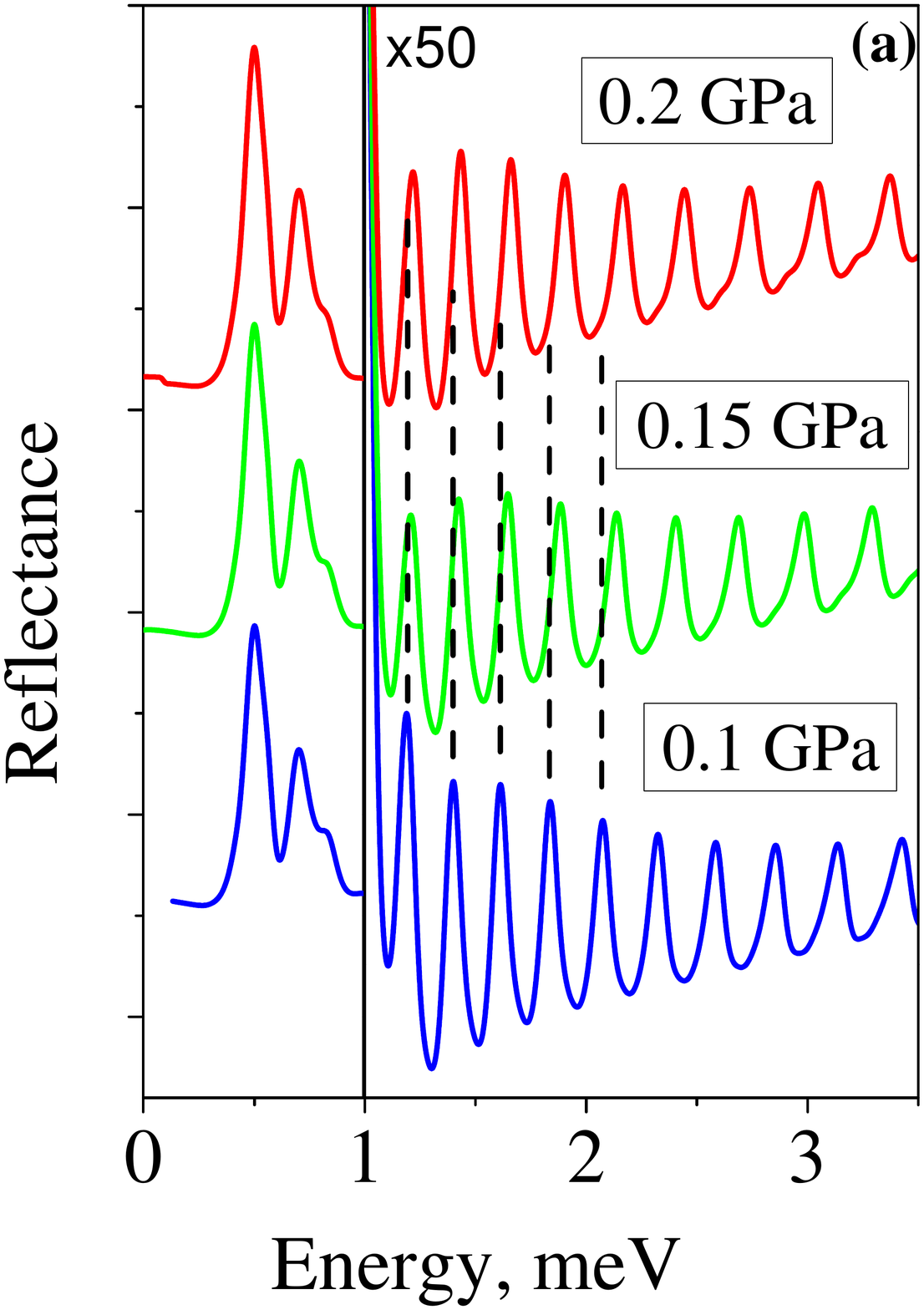}
\includegraphics[clip,width=.49\columnwidth]{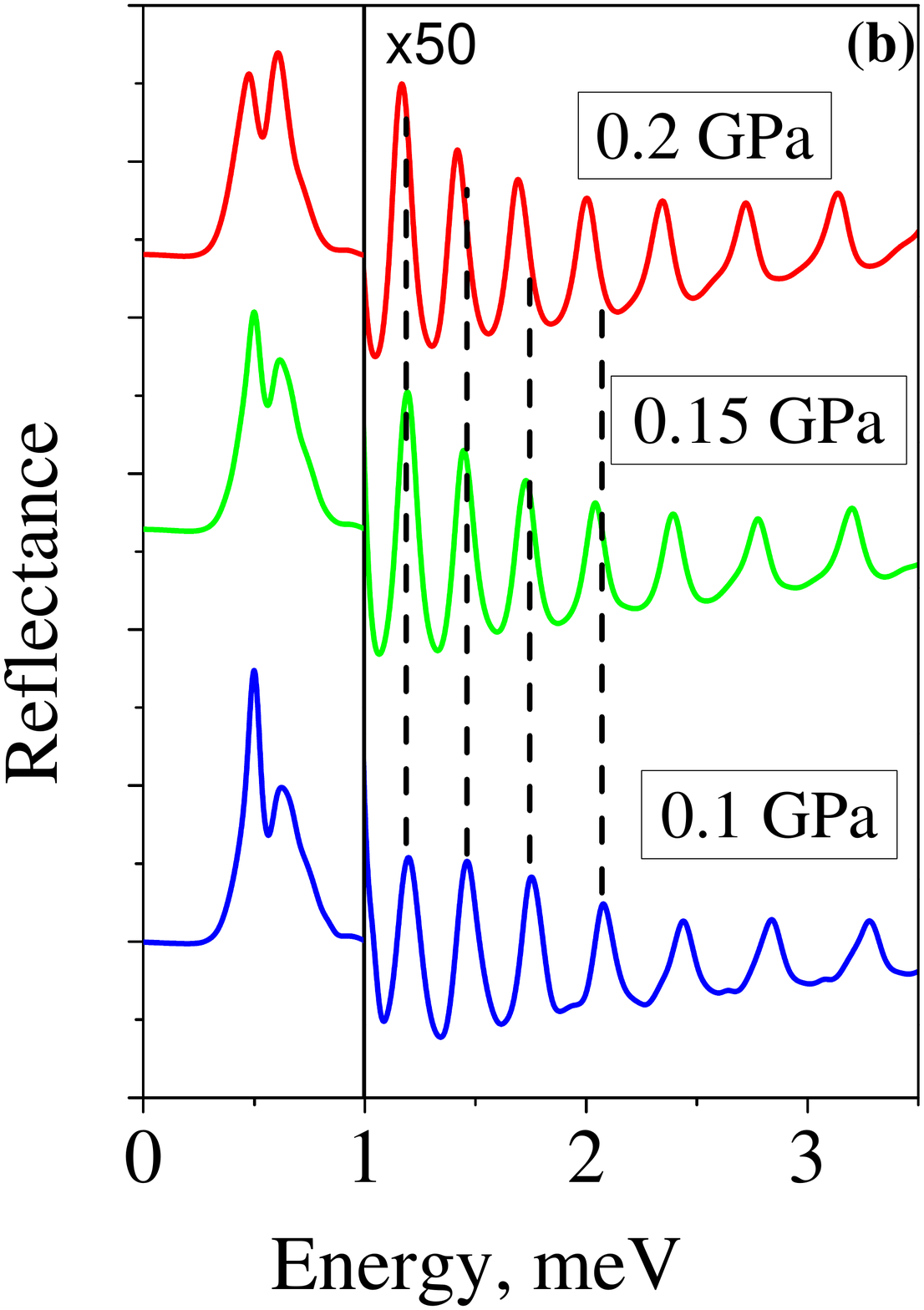}
\caption{Reflection spectra $R^{(++)}(\omega)$ for the GaAs QW with $L_{QW}=700$ nm  in the spectral range of interference of the $h$- and $l$-polariton modes (a and b, respectively). The magnitude of the applied pressure is given near respective curves. The spectra are shifted alond the energy axis to match their dominant features. The energy of dominant features is taken to be zero. If $P=0$, its value $E\approx1515$~meV. The dashed vertical lines allow one to demonstrate the relative shift of the oscillations. The amplitude of spectral oscillation is multiplied by 50.
\vspace{0.5cm}}
\label{fig:HHmass}
\end{figure}

\begin{figure}[t] 
\includegraphics[clip,width=.49\columnwidth]{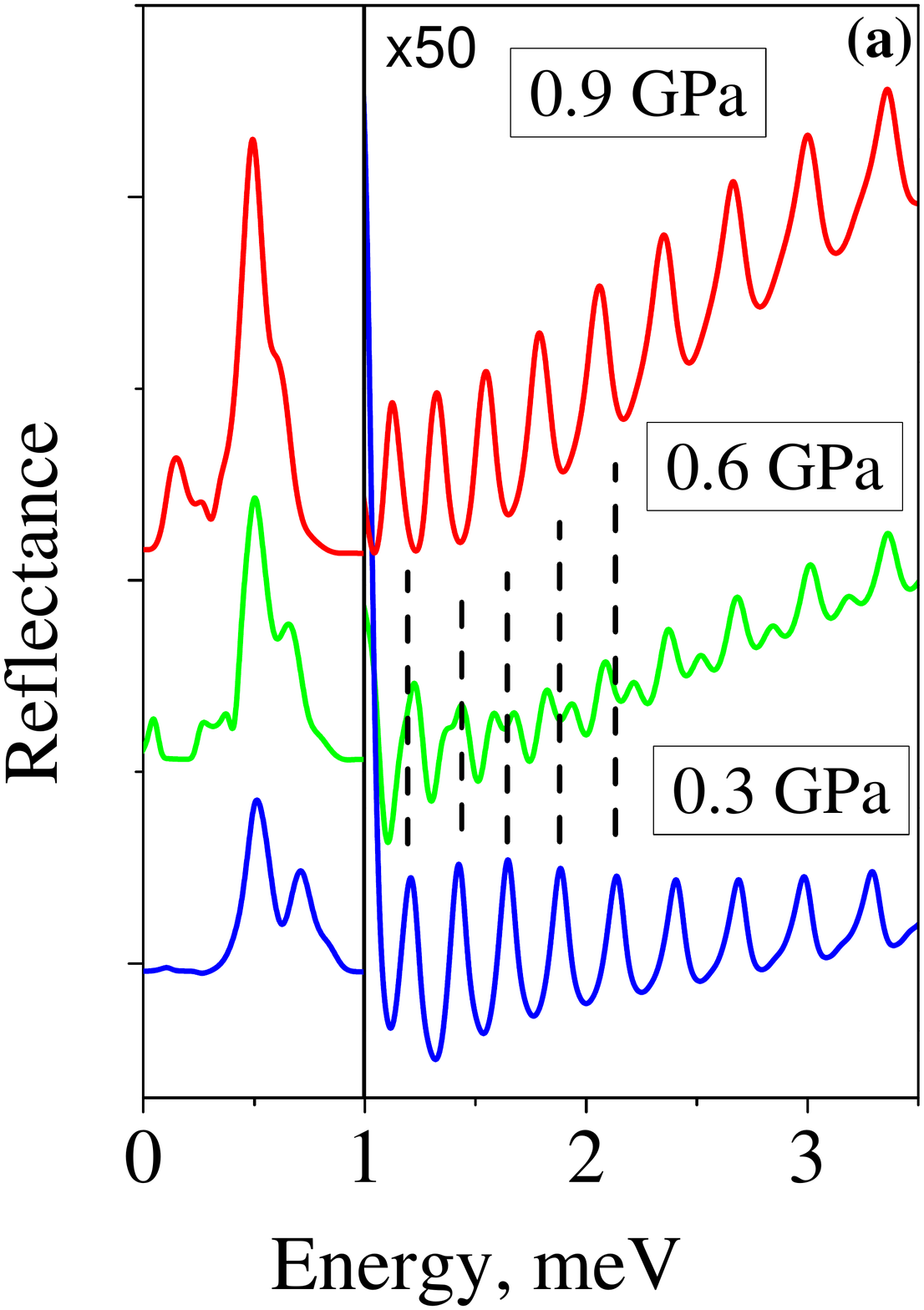}
\includegraphics[clip,width=.49\columnwidth]{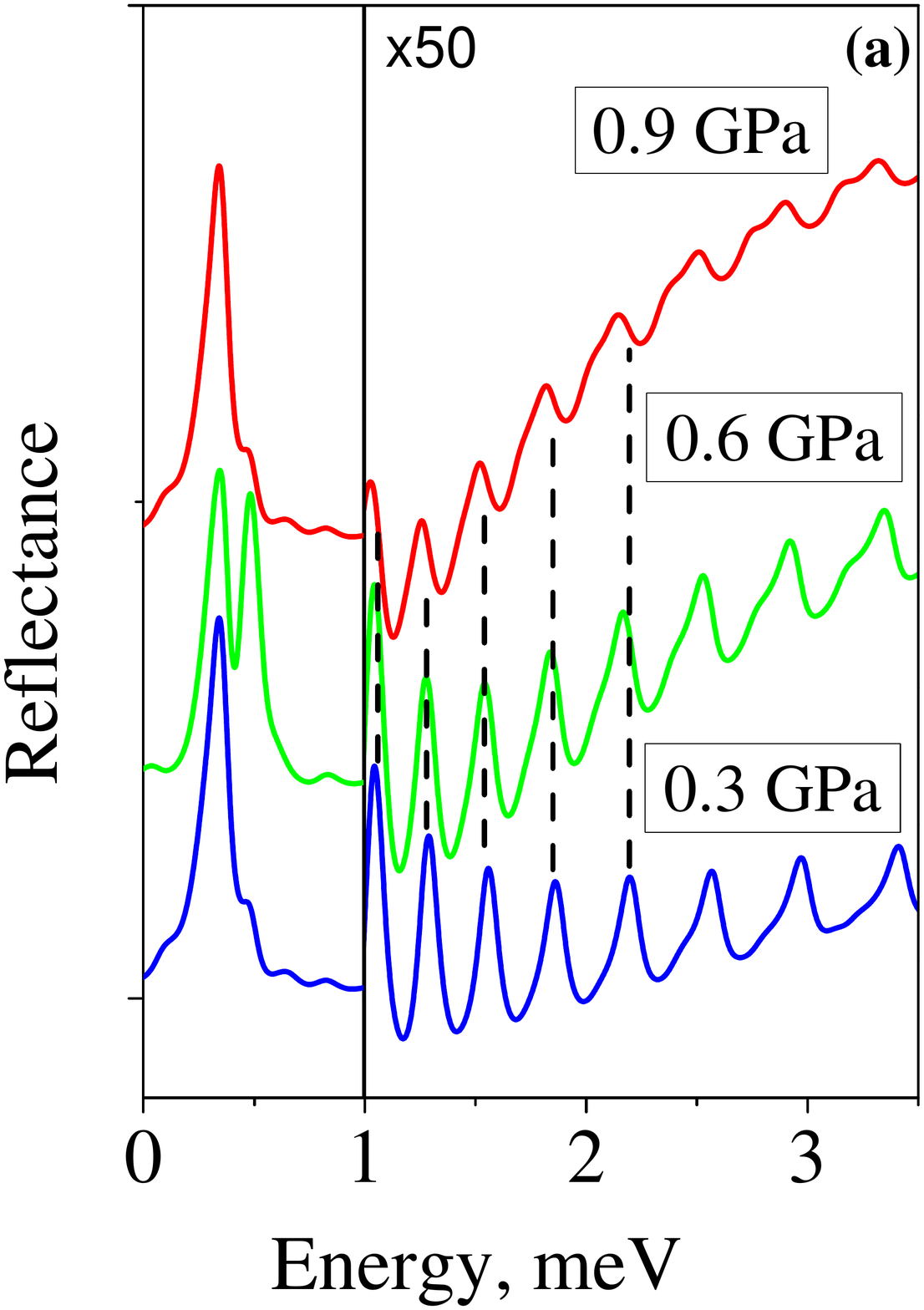}
\caption{Reflection spectra $R^{(++)}(\omega)$ at pressure $P>0.3$ GPa  in spectral range of the $h$- and $l$-polariton modes (a and b, respectively). Other notations are the same as in Fig. \ref{fig:HHmass}. \vspace{0.5cm}}
\label{fig:3}
\end{figure}

\begin{figure}[t] 
\includegraphics[clip,width=.8\columnwidth]{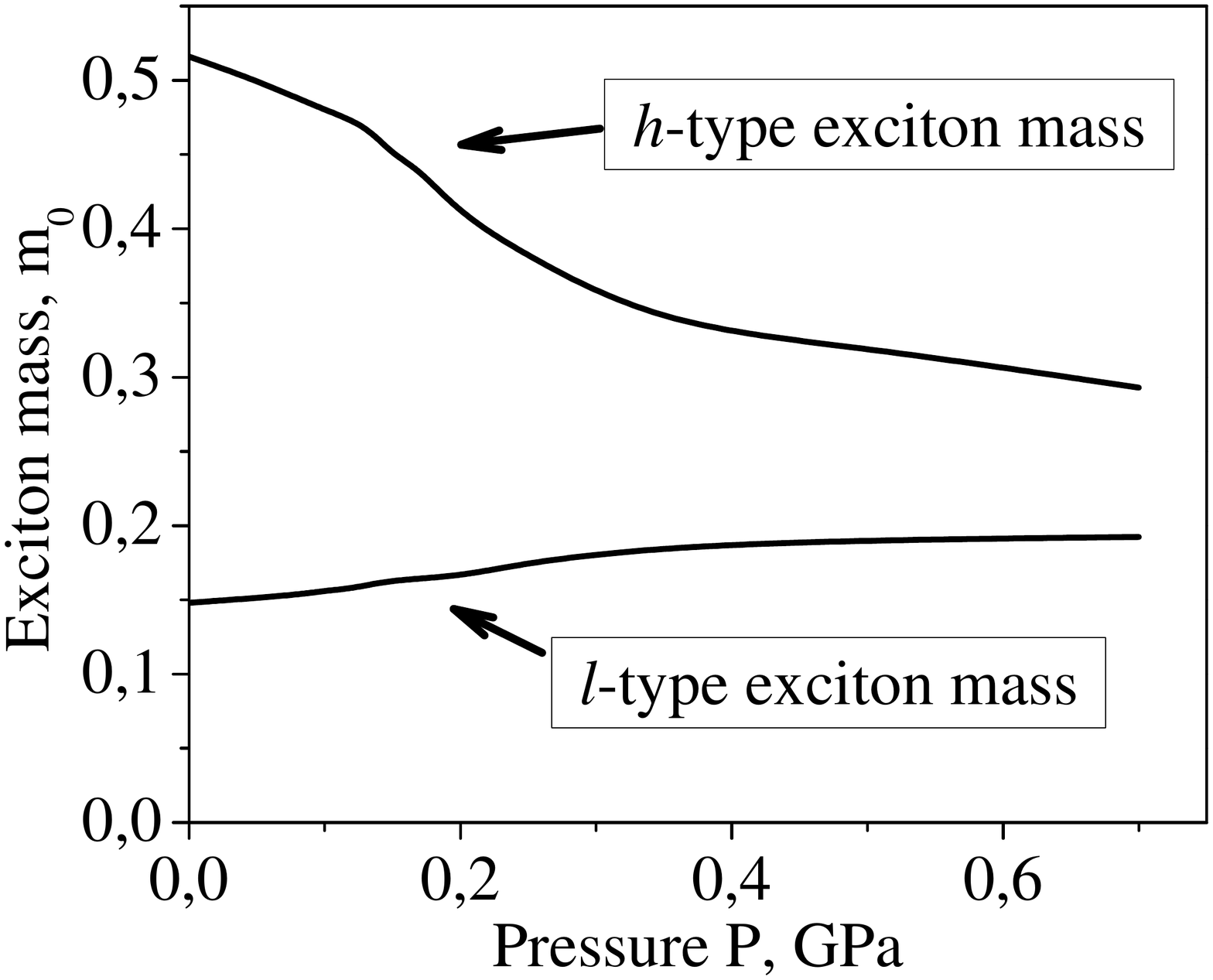}
\caption{Exciton masses in the GaAs crystal as functions of the applied pressure. 
\vspace{0.5cm}}
\label{fig:convergency}
\end{figure}

The co-polarized reflectance spectra calculated in the framework of the described model are shown in Fig.~\ref{fig:HHmass} and Fig.~\ref{fig:3}. Each of the spectrum contains intense peaks and quasi-periodic oscillations. The peaks corresponds to the anti-crossing of the photon and exciton dispersion branches while the oscillations are due to the interference of the exciton-like and photon-like modes  The spectra of l-type  polaritons are strongly shifted to higher energy range at pressure $P > 0.1$~GPa and is not overlapped with those of h-type polaritons. Therefore they can be analyzed separately. As seen in Fig.~\ref{fig:HHmass}a, the distance between the spectral oscillations for $h$-polaritons becomes larger with increasing pressure, which is a result of the aforementioned decrease of the effective mass (see Fig.~\ref{fig:1}). Correspondingly, the spectral oscillations for $l$-polariton become denser  (see Fig. \ref{fig:HHmass}b) due to the increasing mass of the $l$-type excitons. The pressure dependences of masses for $h$- and $l$-excitons  obtained from analysis of the curvature of dispersion branches are shown in Fig.~\ref{fig:convergency}.

At larger pressures, $P\ge0.3$ GPa the "stretching" or "compressing" of the oscillations is no longer observed, see Fig.~\ref{fig:3}.

To discuss this behavior of oscillations qualitatively, we neglect the exciton-photon interaction and calculate the exciton energy from the secular equation with the Hamiltonian  (\ref{eq:hamexfull}):

$$\det|\hat H_X-E\mathbb I|=0.$$
Let us denote $\Theta\equiv(\frac{\hbar^2}{2M_h}-\frac{\hbar^2}{2M_l})$ and
$\Delta\equiv(H_{\varepsilon h}-H_{\varepsilon l})=b(\varepsilon_{xx}+\varepsilon_{yy}-2\varepsilon_{zz})$. Note that the $K$- linear  terms of the Hamiltonian are small in comparison with $\Theta K^2,~V,~\Delta$, so that one may neglect term $A_{3/2}K$ in Hamiltonian Eq.  (\ref{eq:hamexfull}). In this approximation,  the energy of $l$ and $h$ excitons is:
%
$$E_{\pm}=\frac12\Bigl[(H_{h\pm}+H_{l\mp})\pm\sqrt{(H_{h\pm}-H_{l\mp})^2+4V^2}\Bigr]\approx$$
%
$$\frac12\Bigl[(H_{h\pm}+H_{l\mp})\pm\sqrt{(\Theta K^2+\Delta)^2+4V^2}\Bigr],$$
%

Note that quantities $\Delta$ and $V$ linearly depend on the pressure. At small pressure, when $\Theta K^2$ is of the same order as $\Delta$ and $V$, the exciton energy essentially depends on the wave vector, resulting in a relatively fast convergence of exciton masses (see Fig.~\ref{fig:convergency}) and, correspondingly, in stretching and compressing of the $h$ and $l$ oscillations, respectively. At high pressure, $\Delta$ and $V$ are large compared with  $\Theta K^2$ and this term can be neglected. Therefore the convergence of the masses and modification of the oscillations are blocked at high pressures, as can be seen in Figs. ~\ref{fig:convergency} and \ref{fig:3}. 

At pressure $P\ge0.4$ GPa, another effect appears, which is caused by the $K$- linear splitting of dispersion branches already demonstrated in Fig.~\ref{fig:1}. As seen in Fig.~\ref{fig:3}a, the oscillations in the $h$-spectrum decrease in amplitude with increasing pressure. However, at pressure above a certain critical value, $P_{cr}$, the oscillations begin to recover, their phase being opposite to that at low pressures. This phenomenon can be called an ``inversion'' of the oscillation phase. It should be emphasized that the amplitude of the dominant reflection peak is almost independent of the pressure.

Analysis showed that$P_{cr}$ is a function of the QW width, $L_{QW}$. It is approximately inversely proportional to $L_{QW}$ and for the GaAs QW is fitted by dependence: $P_{cr}=a+b/L_{QW}$ with $a=-0.3$ GPa, and $b=480$ GPa/nm. Note that, for $L_{QW}<400$ nm, the critical pressure exceeds the ultimate magnitude for GaAs crystal \cite{Ansp, CardonaPSS}. Nevertheless, since  the spectral oscillations are observable  in the high-quality GaAs QWs with thickness up to 1 micron (see Ref.~\cite{Loginov}), there is real possibility to see this effect experimentally.

\begin{figure}[t] 
\includegraphics[clip,width=.49\columnwidth]{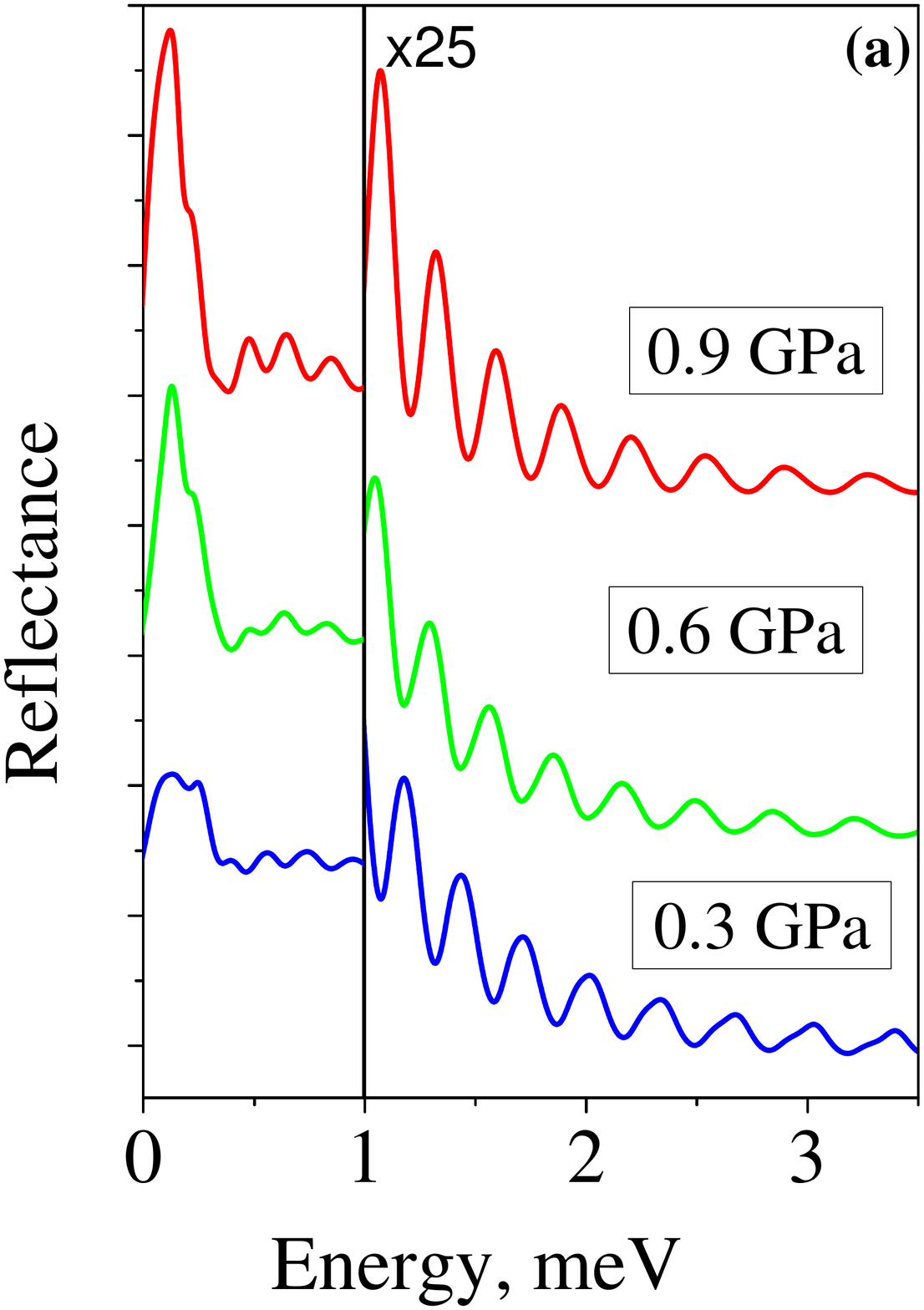}
\includegraphics[clip,width=.49\columnwidth]{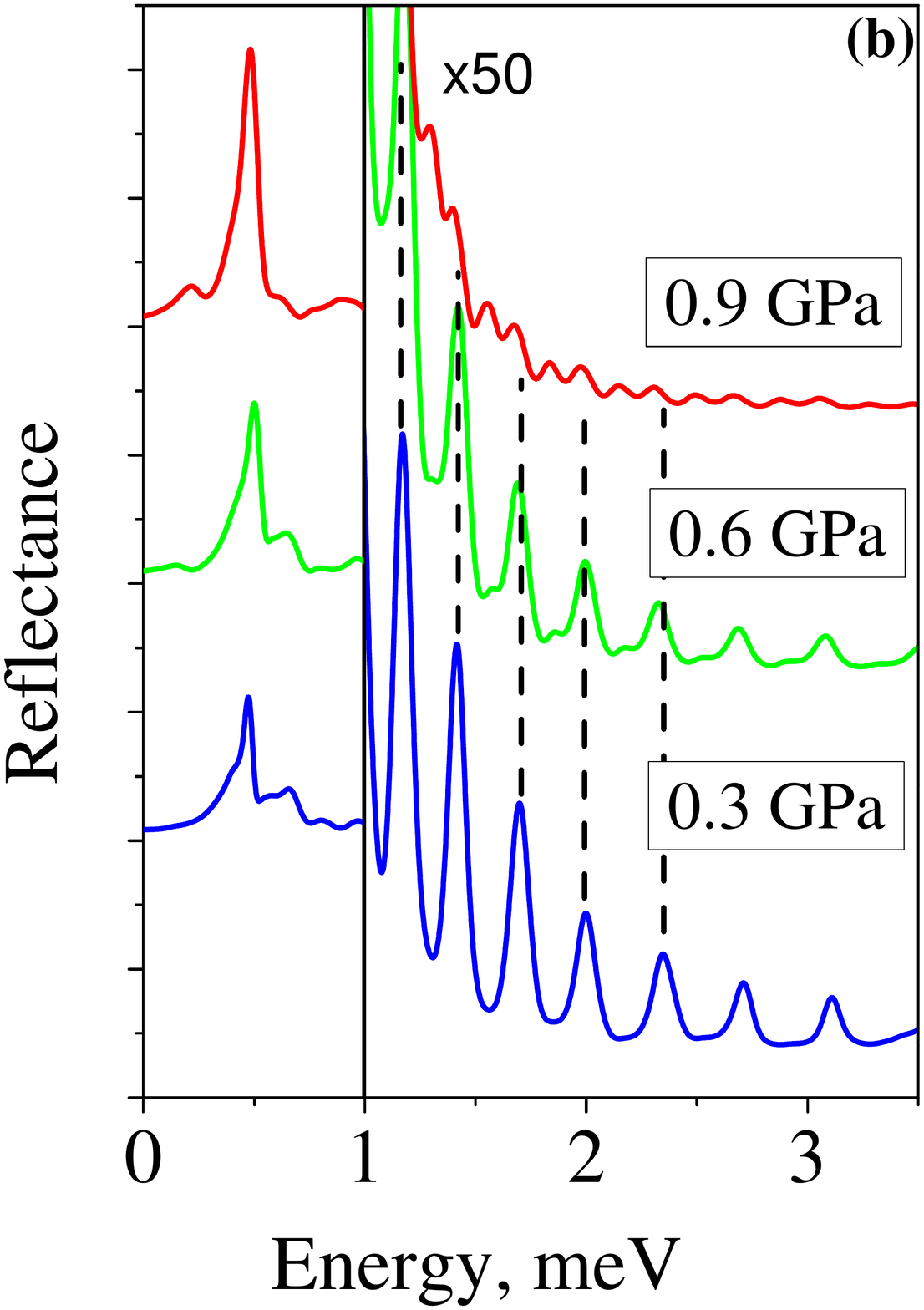}
\caption{Reflection spectra in cross polarization in spectral range of interference of $h$- and $l$-polariton modes (a and b, respectively). Other notations are the same as in Fig. \ref{fig:HHmass}. 
\vspace{0.5cm}}
\label{fig:invspectra300}
\end{figure}

The phase inversion in the  $l$-spectrum appears at higher pressures (see Fig. \ref{fig:3}b).
The reason for that is the absence of $K$-linear splitting for the basic light-hole exciton states  [see Eq.~(\ref{eq:exepsK})]. However, the admixture of the heavy-hole exciton states at high pressures may result in the $K$-linear splitting and, correspondingly, in the inversion of the oscillations phase. A detailed discussion of the origin of phase inversion is presented in the next section.

Let us consider the cross-polarized components of the $h$- and $l$- spectra calculated for the same QW Fig.~\ref{fig:invspectra300}. Their amplitudes are determined by non-diagonal components of permittivity tensor [see Eqs. (\ref{eq:Pol/pm/mp}), (\ref{eq:melnondiog}), and (\ref{eq:conertfield})], which, in turn, is determined by the perturbation $V$, see Eq. (\ref{eq:hamexfull}). 
The amplitude of spectral oscillations rapidly decreases with increasing photon energy in a few meV. 
Note that in co-polarization the oscillations amplitude does not decrease  noticeably in the same spectral range, see Fig. \ref{fig:HHmass}. This difference is due to a rapid divergence of the $h$- and $l$- dispersion branches with the wave vector increase and, correspondingly, to a rapid decrease of mixing of the light-hole and heavy-hole exctions. 

The inversion of the phase of the oscillations is not observed in the cross-polarized $h$ -spectrum, see Fig.~\ref{fig:invspectra300}a. The reason is that these spectra are caused by the strain-induced admixture of the light-hole excitons whose Hamiltonian does not contain $K$-linear terms [see Eq. (\ref{eq:hamexfull})]. On the contrary, the cross-polarized $l$-spectra do reveal the phase inversion effect (Fig. \ref{fig:invspectra300}b) at pressures even lower than that for the co-polarized  $l$-spectra (cf. with Fig. \ref{fig:3}b). This is due to the admixture of heavy-hole excitons, which are $K$- linearly split.

In conclusion of this section, we should note that, even at the maximum possible pressure $P=1$ GPa, the spectral amplitude of co- and cross-polarized oscillations are comparable in magnitude only in a small spectral range of about 0.5 meV above  the exciton transition. At higher energies, the amplitudes of the cross polarized oscillations become negligible small. Thus, the effect of the circular polarizations conversion of incident light is negligible for the rest of spectrum. Therefore, this conversion effect is not considered any more in the next section.

\section{Stress-induced gyrotropy}
\label{sec:dissc}

It follows  from relation (\ref{eq:diogonalsootncirc}) that the uniaxial stress leads not only to birefringence, but also to gyrotropy.
The gyrotropy is due to the $K$-linear splitting of the exciton states with positive and negative projections of the angular momentum on the $Z$ axis. This splitting is described by expression (\ref{eq:exepsK}). It should be emphasized that the necessary (though not sufficient) conditions for the appearance of gyrotropy are the lack of inversion symmetry and the presence of spatial dispersion of excitons  \cite{AgranovichGinzburg}
.

The gyrotropy manifests itself in the appearance of ellipticity of the reflected light at the linearly and circularly polarized incident light. The ellipticity can be described by the ratio of major and minor axes of the polarization ellipse ($E_b$ and $E_s$, respectively) and by the angle $\chi$ 
 between the $X$ axis and the direction of the major ellipse axis (see, e.g., \cite{BornWolf})
. The angle $\chi$ is determined by expression:
\begin{equation}
\chi=\arctan\Bigl[\frac{-{\cal A}+\sqrt{{\cal B}^2-4{\cal AC}}}{2{\cal A}}\Bigr].
\label{eq:chi}
\end{equation}
Here ${\cal A}=\mathrm{Im} E_2\cdot \mathrm{Re}E_1-
\mathrm{Re} E_2\cdot\mathrm{Im}E_1$, ${\cal B}=(\mathrm{Re}E_2)^2-(\mathrm{Im} E_1)^2+
(\mathrm{Im}E_2)^2-(\mathrm{Re}E_1)^2$, ${\cal C}=\mathrm{Im} E_1\cdot \mathrm{Re}E_2-
\mathrm{Re} E_1\cdot\mathrm{Im}E_2$, where $E_1=E^{(+)}_r+E^{(-)}_r$ and $E_2=E^{(+)}_r-E^{(-)}_r$.
The ratio of major and minor axes is:
\begin{equation}
e=\xi\frac{|E_x\cos(\chi)+i\cdot E_y\sin(\chi)|}{|-E_x\sin(\chi)+i\cdot E_y\cos(\chi)|},
\label{eq:ellipsity}
\end{equation}
where $\xi=1$ for $|E_r^{(+)}|/|E_r^{(-)}|>1$ (right-hand elliptical polarization) and $\xi=-1$ for $|E_r^{(+)}|/|E_r^{(-)}|<1$ (left-hand elliptical polarization).

\begin{figure}[t] 
\includegraphics[clip,width=.49\columnwidth]{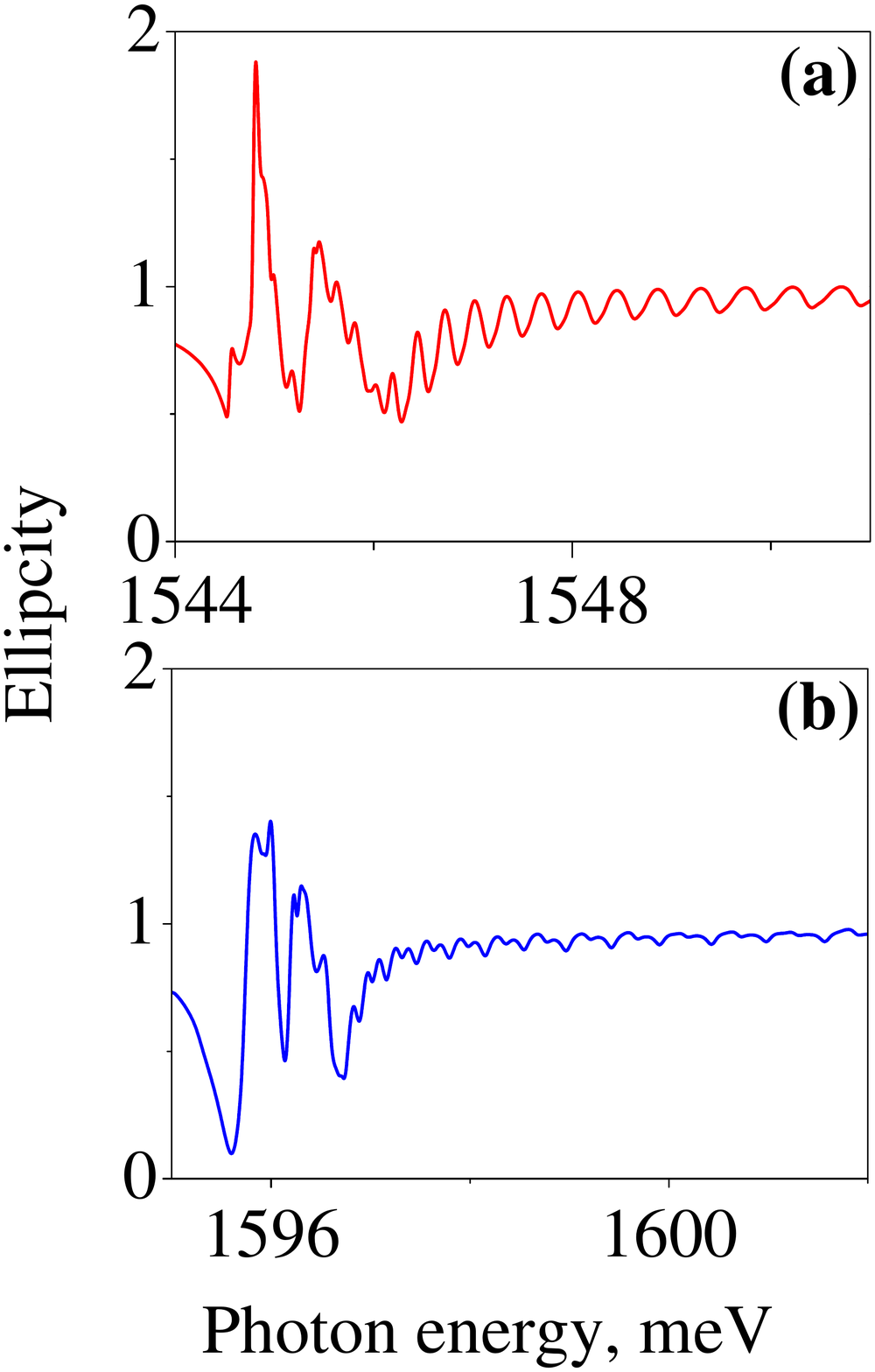}
\includegraphics[clip,width=.49\columnwidth]{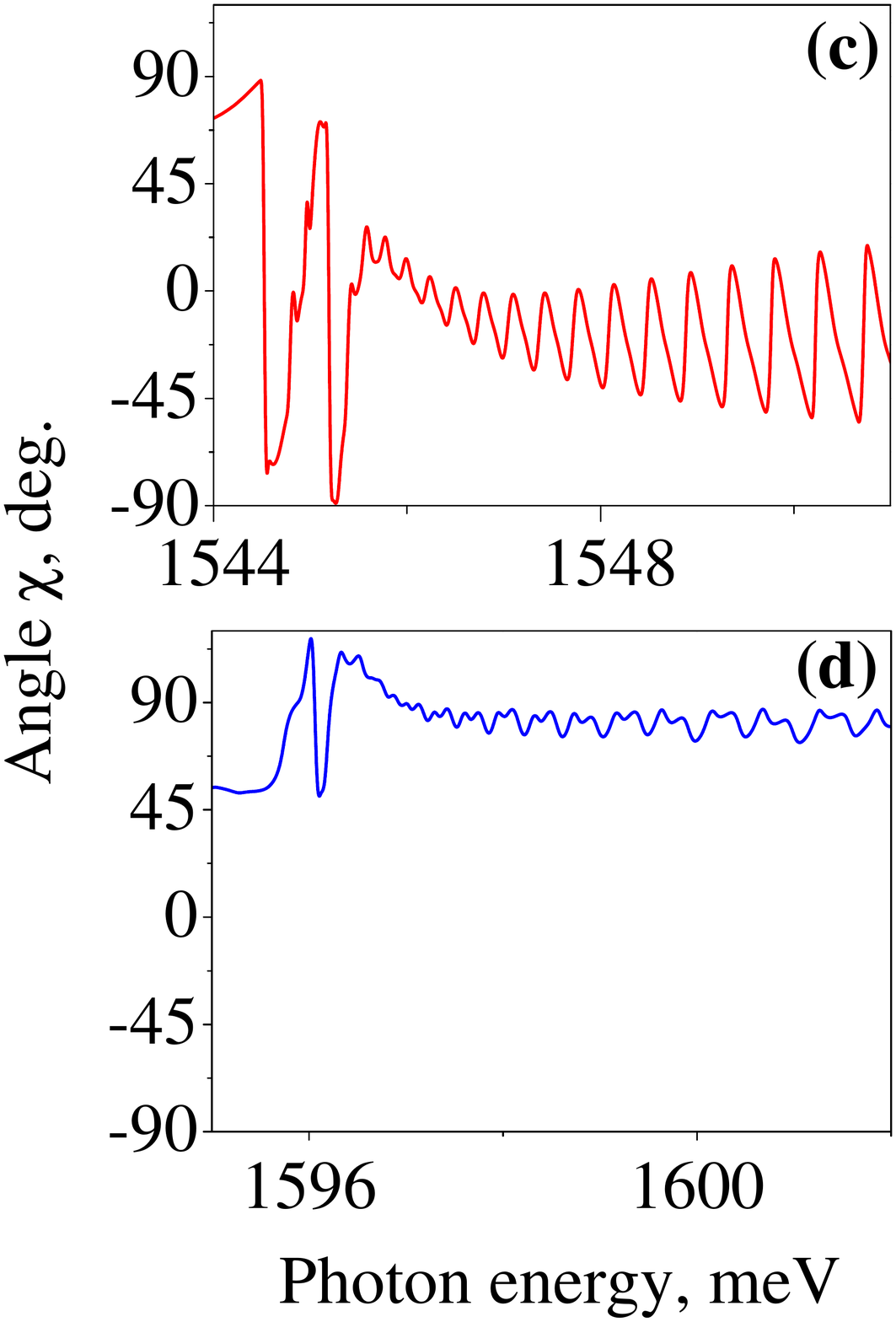}
\caption{The spectra of the  ellipticity, $e(\omega)$, and angle, $\chi(\omega)$, of the reflected light (figures \textbf{a}, \textbf{b}  and \textbf{c}, \textbf{d} respectively). Calculations are done for the GaAs QW with $ L_ {QW} = 700 $ nm at $P=0.8$ GPa for the spectral range of $h$- and $l$-polariton modes (rad and blue curves, respectively). The incident light has the right circularly polarization.
\vspace{0.5cm}}
\label{fig:Ellipcity}
\end{figure}

We have calculated the spectra of $e(\omega)$ and $\chi(\omega)$ for $h$- and $l$-polaritons at pressure $P=0.8$ GPa (Fig.~\ref{fig:Ellipcity}). As seen, the spectrum of $e(\omega)$ consists of a set of oscillations superimposed on the smoothly varying background. The magnitude of $e(\omega)$ significantly differ from unity only in the range of the anti-crossing of exciton and photon dispersion curves. This is caused by the strong mixing of the photon-like and exciton-like modes in this range, which leads to a $K$-linear splitting of the photon-like branch. The ellipticity above the anticrossing is caused by the $K$-linear splitting only of the exciton-like branches. 

Quantity  $e(\omega)$ weakly oscillates in this spectral range about $+1$. The angle $\chi$ oscillates about zero for the $h$-polariton and about $\pi/2$ for the $l$-polariton. This means that the major axes of these excitons are perpendicular to each other. They swing about the direction of applied pressure (for the $h$-polariton) and perpendicular to it (for the $l$-polariton) as the light frequency is varied.  The period of these oscillations coincides with that in the reflection spectra. 

\section{Discussion}

In this section we discuss in more detail specific mechanisms of the phase inversion of spectral oscillations. For simplicity, we consider the light-hole exciton, we restrict ourselves by the consideration of the heavy-hole exciton only. In this case, the Maxwell's boundary conditions and the Pekar's ABC are reduced to a simpler form described in Ref. \cite{Pekar}. We illustrate the mechanism of the phase inversion by calculations of the reflection spectra in framework of the multipath interference model described, e.g., in Ref. \cite{BornWolf}. 

Let a circularly polarized light wave $E_i$  falls onto the left boundary of a QW. This wave is partially reflected  (wave $E_r^{(0)}$) and partially penetrates into the QW. In the QW layer, the exciton-like ($ E_x $) and photon-like ($ E_p $) polariton waves propagate in the forward direction along the Z axis. Amplitudes of waves  $E_r^{(0)}$, $E_x$, and $E_p$ are determined from the boundary conditions  (\ref{system0}). 
When the exciton-like wave $E_x$ reaches the right interface of QW, it partially penetrates to the right barrier (wave $E^t_x$) and partially is reflected. After this reflection, already two reflected waves, exciton-like wave ($E_ {xx}$) and the photon-like wave ($E_{xp}$), propagate in the backward direction. Amplitudes of waves $E^t_x$, $E_ {xx}$, and $E_{xp}$ are also determined from the boundary conditions at the right interface. Similar processes occur with the photon-like wave $E_p$ at the right boundary. Thus, four waves propagate from the right interface of the QW in the negative direction: $E_{xx}$, $E_{xp}$, $E_{pp}$ and $E_{px}$. Similarly, the waves $E_{xx}$, $E_{xp}$, $E_{pp}$ and $E_{px}$ can partially penetrate into the left barrier and partially be reflected from the left barrier into the QW. After this reflection, eight waves propagate in the QW layer in the forward direction. These waves start a new cycle, as described above. 

Thus, an infinite number of waves are generated in the QW during propagation of light. The electric field amplitudes of these waves can be expressed through  $E_i$ 
if the amplitude coefficients of reflection and transmission are known. However, it is technically difficult to sum all the possible contributions to the reflection because of the large number of waves created at each interface. However, our calculations show that, to obtain a satisfactory agreement with the results obtained by the transfer matrix method, it is enough to summarize a few main contributions into the reflected wave. Waves with similar subscripts can be summarized as an infinite geometric sequence. This allows to find $E_ {rX}$, $E_ {rxpX}$, $E_ {rxpP}$, $E_ {rpxX}$, and $E_{rpxP}$, where capital subscripts indicate summation of an infinite number of similar waves, e.g., $E_{rxpX} = E_{rxpxx} + E_{rxpxxxx} + \ldots$.
\begin{figure}
\includegraphics[clip,width=.99\columnwidth]{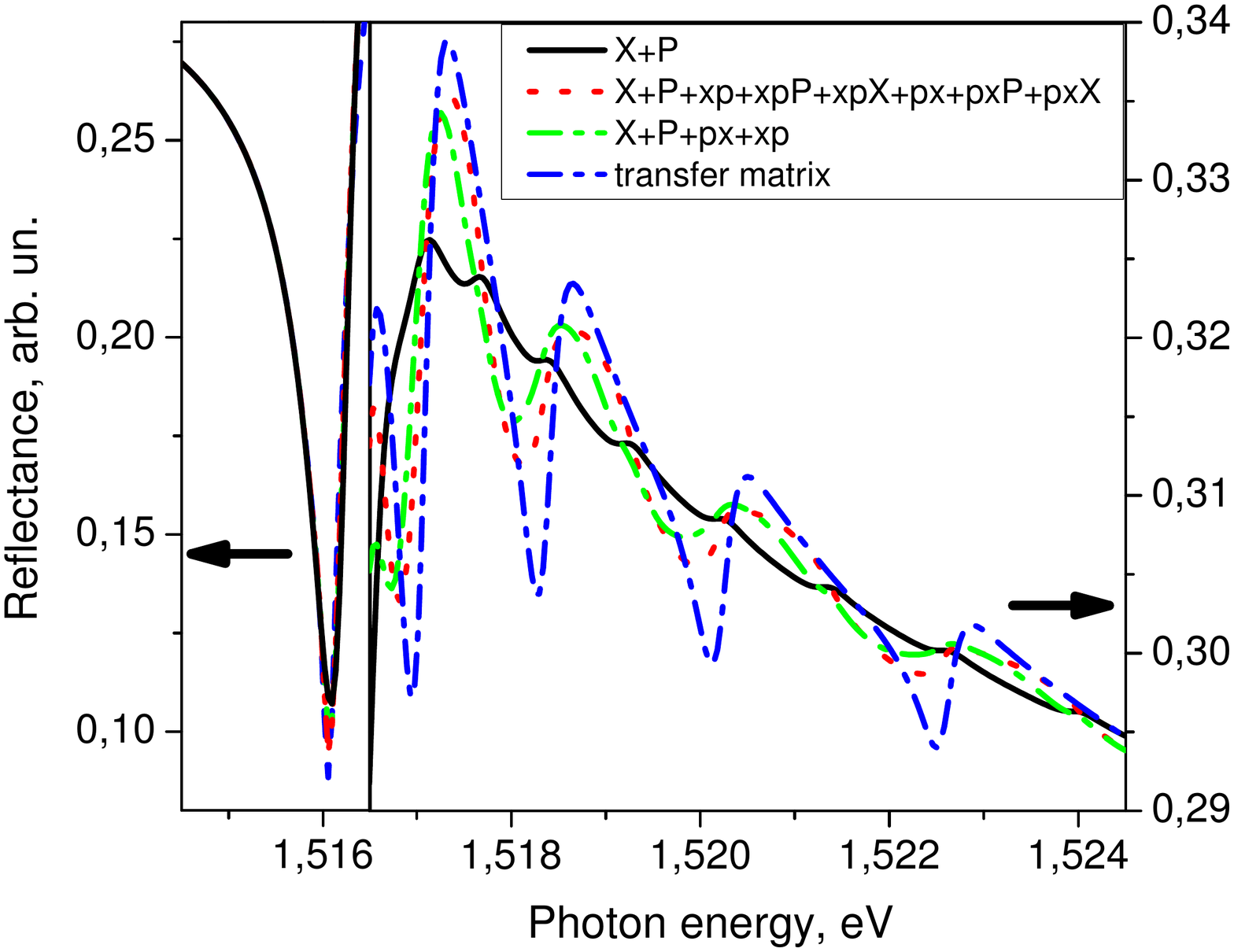}
\label{3refl_P0}
\caption{Reflection spectra calculated at $P=0$. Solid curve is $|E_{r0} + E_{rP} + E_{rX}|^2$, rare dotted curve is $|E_{r0} + E_{rP} + E_{rX}+E_{rxp}+E_{rpx}+E_{rxpP}
+E_{rxpX}+E_{rpxP}+E_{rpxX}|^2$, dash-dot curve is $|E_{r0}+E_{rX}+E_{rP} +E_{rxp}+E_{rpx}|^2$, quick dotted line - calculation method of transfer matrices. The amplitude of incident wave $|E_i|=1$.}
\end{figure}

Results of the calculation are shown in Fig.~\ref{3refl_P0} for pressure $P=0$. If only photonic waves, $E_ {rP}$, and excitonic waves, $E_{rX}$.are taken into account, the calculated spectrum contains the main peak and almost non-oscillating background (solid line in the figure). Interference of waves $E_{rxp}$ and $E_{rpx}$ provides polariton oscillations, which spectral positions coincide with those calculated by the transfer matrix method. Hence,  the spectral oscillations are the result of the interference of the polaritonic waves, which propagate as photon-like waves in one direction and as exciton-like waves in the opposite direction, that is $E_{rxp}$ or $E_{rpx}$. That is the reason why the energy distance between the oscillation peaks is approximately twice larger then the distance between the neighboring energy levels of the exciton size quantization \cite{Loginov}. Extending the consideration to the waves $E_{rxpP}$, $E_{rxpX}$, $E_{rpxP}$, and $E_{rpxX}$ improves agreement with the exact calculation, but does not provide any additional spectral features.

Upon application of pressure, the waves $E_{rxp}$ and $E_{rpx}$ are no longer equivalent: their amplitudes remain roughly the same, but the phases are different. 
Let us denote $\Delta K_x=(K_{x+}+K_{x-})/2,~K_{x0}=(K_{x+}-K_{x-})/2$ and $K_{p+}\approx-K_{p-}\approx K_{p0}$, where $K_{x0}$ and $K_{p0}$ are the wave vectors at zero pressure. Note that wave vectors $ K_ {x-} $ and $ K_ {p-} $ are negative. The amplitude reflection coefficient with taking into account the major contributions is:
\begin{widetext}
$$r = \frac{E_{r0} + E_{rP} + E_{rX} + E_{rxp} + E_{rpx}}{E_i} = r_{00} +\frac{t_{p0}^{-}r_{pp}^{+-}t_{0p}^{+}e^{i\left( K_{p-}+K_{p+} \right)L_{QW}} }{1-r_{pp}^{-+}r_{pp}^{+-}e^{i\left( K_{p-}+K_{p+} \right)L_{QW}}} +$$
\begin{equation}
+ \frac{t_{x0}^{-}r_{xx}^{+-}t_{0x}^{+}e^{i\left( K_{x-}+K_{x+} \right)L_{QW}} }{1-r_{xx}^{-+}r_{xx}^{+-}e^{i\left( K_{x-}+K_{x+} \right)L_{QW}}} +
t_{p0}^{-}r_{xp}^{+-}t_{0x}^{+}e^{i\left(K_{x+} + K_{p-} \right)L_{QW}} + t_{x0}^{-}r_{px}^{+-}t_{0p}^{+}e^{i\left(K_{x-} + K_{p+} \right)L_{QW}} \approx 
\label{cos}
\end{equation}
  $$\approx r_{00} + \frac{A_P e^{2i K_{p0}L_{QW}}}{1-B_P e^{2i K_{p0}L_{QW}}} + \frac{A_x e^{2i K_{x0}L_{QW}}}{1-B_X e^{2i K_{x0}L_{QW}}} + 2A_{xp}e^{i(K_{x0} + K_{p0})L_{QW}} \cos{\frac{\Delta K_x L_{QW}}{2}}$$
\end{widetext}

where conversion coefficients $r_{00}, t_{0p}^{+},t_{0x}^{+}$ are given in the Appendix B [see Eq. (\ref{eq:inftyindex})], and the other coefficients are calculated in a similar way.

Equation (\ref{cos}) allows one to understand the effect of the inversion of the oscillation phase. When the last term is zero, that $\cos(\Delta K_xL_{QW}/2)=0$ , the oscillations disappear. It follows from that the critical pressure is: 
$$P_{cr}=\frac{\pi\hbar^2}{2m_h(|j|C_6+|j|^3C_8)(S_{11}-S_{12})L_{QW}},$$
where $j=3/2$, and other notations are the same as in Eqs. (\ref{eq:linearhole}) and (\ref{eq:epsilonP}). Below this pressure, $\cos(\Delta K_xL_{QW}/2)>0$, the last term in Eq. (\ref{cos}) is positive and the oscillation phase has the same sign. Above it, $\cos(\Delta K_xL_{QW}/2)<0$, and the sign is opposite.

\section{Conclusion}

We developed a theory of the interference of polariton modes in a heterostructure with a wide quantum well, subject to uniaxial stress perpendicular to the growth axis. The model includes the photon-exciton interaction and the strain-induced effects described by the Bir-Pikus Hamiltonian. In particular, the $K$-linear terms appearing in the hole Hamiltonian due to the strain are taken into account.

The first one is the convergence of masses of the heavy-hole and light-hole excitons with increasing pressure. The analysis shows that this effect is due to mixing of these excitons, which is described by the Hamiltonian of Bir and Pikus.

Another effect of deformation is the suppression of oscillations in the spectra of the circularly co-polarized reflection at some critical pressure and their recovery when the pressure increases further. This effect is accompanied by the inversion of oscillation phase. The phenomenon is a direct consequence of a more general effect of the $K$-linear splitting of the valence band $\Gamma_8$ that is induced in crystals without the inversion symmetry by a uniaxial stress.

The analysis shows that, in spectra of the circularly cross-polarized reflection, the spectral oscillations are also could be observed. The amplitude of these oscillations increases with the increasing pressure. However, even at the highest possible pressure, the amplitude of these oscillations is much smaller than that of oscillations in the co-polarized reflection. The effect of phase inversion for light-hole excitons is also could be observed due to the mixing with the heavy-hole excitons, but it occurs at higher pressures.

The estimates made in this paper show that the considered effects can be experimentally observed in heterostructures with relatively wide GaAs/AlGaAs quantum wells at sub-critical pressures $P<1$ GPa, at which the crystal is not yet damaged

\section*{Acknowledgments}
%

We appreciate valuable discussions with I. Ya. Gerlovin, M. M. Glazov, A.V. Sel'kin and A. V. Koudinov. This work was partially supported by the Russian Ministry of Education and Science (Contract No. 11.G34.31.0067 with SPbSU). The authors acknowledge Saint-Petersburg State University for a research grant 11.38.213.2014.

\section*{Appendix A}
\label{sec:A}
According to Eqs.~(\ref{eq:linearelectron}) and (\ref{eq:linearhole}), the uniaxial stress gives rise to linear in $k$ terms in the Hamiltonians of both electrons and holes. As a consequence, the terms linearly dependent on the wave vector $K$ appear in the exciton Hamiltonian. To find these terms, one should go from operators $\hat k_\alpha$ to the operators $\hat K$ and $\hat p_\alpha$ by means of substitution of expressions  (\ref{eq:tranzoper}) in Eqs. (\ref{eq:linearelectron}) and (\ref{eq:linearhole}). This substitution gives rise to many terms in exciton Hamiltonian. We consider only the terms linearly dependent on $ \hat K_z $. Other terms contain operators  $ \hat p_x, ~ \hat p_y, ~ \hat p_z $ , which mix 1s- and np-  excitonic states and lead to a shift of the excitonic spectrum as a whole. Since this shift is much less than the shift of excitonic spectrum described by the Bir-Pikus Hamiltonian (\ref{eq:BirPikusfull}), we ignore it in the further analysis.

Among the terms linearly dependent of $\hat k_z$, the first and the last terms in Eq.~(\ref{eq:linearhole}) can be excluded from consideration because $\hat k_z$ is present only in terms containing  $J_x$, $J_y$ and $V_z$. These terms describe the mixing of the light-hole and heavy-hole excitons, whose strength is inversely proportional to the splitting of these excitons described by  $H_\varepsilon$ [see Eq.~(\ref{eq:BirPikusfull})]. For the characteristic magnitudes of the exciton wave vector $K$ considered in our work, $H_\varepsilon\gg H_v^{(\varepsilon k)}$, therefore such mixing has little effect on the energy of exciton, and we neglect it. We also do not consider operator described by  Eq. (\ref{eq:linearelectron}), since constants $C_3$ and $C'_3$ are much less than $C_6$ and $C_8$ entering Eq. (\ref{eq:linearhole}). 

Besides the above terms, there are $k$-linear contributions to the hole Hamiltonian which are independent of strain \cite{PikusMaruschak}. These terms mix states of the heavy and light holes, as well as the ground and excited states of exciton. Our analysis shows that they do not result in $k$-linear splitting of 1s-exciton states. Material constants determining the magnitude of the mixing are small for most crystals, therefore these terms are not considered in present work.

\section*{Appendix B}

The electric field amplitudes,  $E_i$, $E_ {ri}$, $E_x$, and $E_p$, are related to each other by the Maxwell's and Pekar's boundary conditions:

\begin{equation}
\begin{array}{c}
E_i + E_{ri} = E_x + E_p \\
n_0E_i - n_0E_{ri} = n_{x+}E_x + n_{p+}E_p \\
\chi (K_{x+}, \omega)E_x + \chi(K_{p+}, \omega)E_p = 0,
\end{array}
\label{system0}
\end{equation}
$$\text{where  } \chi(K_{p,x+})=\frac{\epsilon_0\hbar^2\omega_{LT}}{H_{Xh}+H_{\epsilon h}+A_{3/2}K_{p,x\pm}}.$$

From these conditions, we find the amplitude coefficients for transmission and reflection:

\begin{equation}
\begin{array}{c}
r_{00} = \frac{E_{ri}}{E_i} = \frac{-n_{p+}+\alpha_{+}^{+} n_{x+}+(1-\alpha_{+}^{+})n_0}{n_{p+}-\alpha_{+}^{+} n_{x+}+(1-\alpha_{+}^{+})n_0}, \\
t_{0p}^{+} = \frac{E_p}{E_i} = \frac{2n_0}{n_{p+}-\alpha_{+}^{+} n_{x+}+(1-\alpha_{+}^{+})n_0} ,\\
t_{0x}^{+} = \frac{E_x}{E_i} = \frac{-2\alpha_{+}^{+} n_0}{n_{p+}-\alpha_{+}^{+} n_{x+}+(1-\alpha_{+}^{+})n_0},
\end{array}
\label{eq:inftyindex}
\end{equation}
\begin{center}
where $\alpha_{+}^{+} = \frac{\chi(K_{p+}, \omega)}{\chi(K_{x+},\omega)}$
\end{center}
Here, to simplify the notation of coefficients, the plus and minus signs indicating the direction of the wave propagation are transferred from the subscripts to the superscripts.

\end{document}